\documentclass[twocolumn]{aa}
\usepackage{graphicx}
\usepackage{physics,amsmath}
\usepackage{multirow}
\usepackage{txfonts}

\usepackage{natbib}
\bibpunct{(}{)}{;}{a}{}{,} 
\usepackage{xcolor}
\usepackage{caption}
\usepackage{subcaption}
\usepackage[colorlinks=true, allcolors = blue]{hyperref}

\newcommand{\fix}[1]{{#1}}

\begin{document} 
   \title{Structure formation in O-type stars and Wolf-Rayet stars}
   \author{
          C. Van der Sijpt\inst{\ref{inst:KUL}},
          J. O. Sundqvist\inst{\ref{inst:KUL}}, D. Debnath\inst{\ref{inst:KUL}}, F. A.  Driessen\inst{\ref{inst:amsterdam}} and N. Moens\inst{\ref{inst:KUL}}
          }

   \institute{{Institute of Astronomy, KU Leuven, Celestijnenlaan 200D, 
              3001, Leuven, Belgium \label{inst:KUL}\\
              \email{cassandra.vandersijpt@kuleuven.be}}
    \and 
        {Anton Pannekoek Institute for Astronomy, University of Amsterdam, 1090GE Amsterdam, The Netherlands \label{inst:amsterdam}}}

   \date{Received 26/09/2024; accepted 10/01/2025}
   \abstract 
   % context
   {Turbulent small-scale structures in the envelopes and winds of massive stars have long been suggested as the cause for excessive line broadening in the spectra of these stars that could not be explained by other mechanisms such as thermal broadening. However, while these structures are also seen in recent radiation-hydrodynamical simulations, their origin, particularly in the envelope, has not been extensively studied.}
   % aims
   {We study the origin of structures seen in 2D radiation-hydrodynamical unified stellar atmosphere and wind simulations of \fix{O stars} and Wolf-Rayet stars. Particularly, we study whether the structure growth in the simulations is consistent with sub-surface convection, as is commonly assumed to be the origin of this turbulence.}
   % methods
    {Using a linear stability analysis of the optically thick, radiation-pressure dominated envelopes of massive stars, we \fix{identified} multiple instabilities that could be driving structure growth. We \fix{quantified} the structure growth in the non-linear simulations of \fix{O stars} and Wolf-Rayet stars by computing density power spectra and tracking their temporal evolution. Then, we \fix{compared} these results to the analytical results from the stability analysis to distinguish between the different instabilities.}
    % results
    {The stability analysis leads to two possible instabilities:
    %the possible instabilities. Two possible instabilities were found: 
    the convective instability and an acoustic instability that is a local variant of so-called strange modes. Analytic expressions for the growth rates of these different instabilities are found.
    %the so-called `strange mode instability'. 
    %It is found that the growth rates $\omega$ for these instabilities have a distinct dependence on wavenumber $k$: $\omega \sim 1/k^2$ for the convective instability and $\omega \sim k^0$ for the acoustic instability. 
    %JS-suggested re-formulation in view of recent results. 
    In particular, strong radiative diffusion damps the growth rate  
    $\omega$ of the convective instability in this regime leading to a distinct $\omega \sim 1/k^2$ dependence on wavenumber $k$. From our power spectra analysis of the simulations, however, we find that structure growth rather increases with $k$ -- tentatively as $\omega \sim \sqrt{k}$. }   
    %for the \fix{O stars} and $\omega \sim k^0$ for the Wolf-Rayet stars which means that neither are compatible with the convective instability as a formation mechanism and the Wolf-Rayet stars could be compatible with the acoustic instability.
    % conclusion
    {Our results suggest that, contrary to what is commonly assumed, structures in luminous O and Wolf-Rayet star envelopes do not primarily develop from the sub-surface convective instability. Rather the growth seems compatible with either the acoustic instability in the radiation-dominated regime or with Rayleigh-Taylor type instabilities, although the exact origin remains inconclusive for now.} 

   \titlerunning{Structure formation in O-type stars and Wolf-Rayet stars}
   \authorrunning{C. Van der Sijpt et al.}
   
   \maketitle

    \section{Introduction}
    \label{introduction}

    Generally, massive stars are believed to have convective cores and smooth radiative envelopes. The current unified model atmosphere codes, covering the sub-surface layers up to the wind outflow simultaneously, assume the atmosphere to be spherically symmetric, stationary, and in radiative equilibrium in order to model it in 1D (see for example, \textsc{FASTWIND}; \citealt{santolaya1997,puls2005,sundqvist2018,puls2020}, \textsc{CMFGEN} \citealt{hillier1998,hillier2001}, \textsc{POWR}; \citealt{grafener2002,hamann2004,sander2012}). Nonetheless, there have been observational and theoretical indications showing that the assumption of a smooth envelope and wind may not hold. \\
    \\
    Particularly, excessive line broadening in \fix{O stars} has been linked to "macroturbulence," large-scale motions of clumps near the surface \citep{conti1977,howarth1997,simondiaz2017}. Furthermore, the winds of these stars are believed to be clumped as well \citep{eversberg1998,puls2006,oskinova2007,sundqvist2010,surlan2012,hawcroft2021,rubiodiez2021,brands2022}. This implies that both the envelope and the wind of these \fix{O stars} are structured instead of smooth. Also for classical Wolf-Rayet (WR) stars, which are believed to be evolved \fix{O stars}, their envelopes and winds \fix{were} shown to be structured \citep{moffat1988,lepine1994,dessart2002,crowther2007,michaux2014}. Moreover, recent multi-dimensional modeling efforts \citep{jiang2015,moens2022,debnath2024} indeed show many structures in the atmospheres of both \fix{O stars} and WR stars. However, while the structure formation mechanism in the wind (the line-deshadowing instability) has been extensively studied \citep{owocki1984,owocki1985,owocki1991,rybicki1990,owocki1996,sunqvist2018b}, the origin of the structures in the near-surface layers is significantly less studied.
    Since \citet{opal} discovered an increase in opacity due to recombination of iron group elements around $\sim 200$ kK, the so-called iron opacity peak or iron bump, structures in these near-surface layers have usually been attributed to a sub-surface convection zone triggered by this increase in opacity \citep{cantiello,jiang2015}. However, a linear stability analysis by \citet{bs03} showed that optically thick radiation-dominated gases are prone to multiple instabilities, and not only the convective instability. Thus far, it has not been formally studied which instability actually triggers the turbulence in the envelopes of these massive stars. The turbulent properties of these envelopes, which might be dependent on the formation mechanism, impact the emergent stellar structure (see, for example, the discussion in \citealt{debnath2024} on turbulent pressure and convective energy transport). These turbulent properties are also important regarding recent efforts to encode multi-dimensional effects in 1D model atmosphere codes \citep{gonzalez2024}. Therefore, it is important to understand the physics behind the structure formation. \\
    \\
    %Based on 
    Building on the work by \citet{bs03}, we study possible instabilities in the envelopes of massive stars and compare theoretical predictions to the structure growth in multi-D simulations of WR stars and \fix{O stars} from \citet{moens2022} and \citet{debnath2024}, respectively. By studying and comparing these different simulations, we probe how the structure formation mechanism may change with the different conditions that prevail in WR and \fix{O star} envelopes. We particularly focus on formally showing whether the convective instability is indeed driving structure formation in these massive stars.
    %Eddington factor or with structural differences due to a different wind launching mechanism 
    \\
    \\
    In Section \ref{background}, we give a summary of the radiation-hydrodynamics equations that are solved and the numerical simulation methods used in \citet{moens2022} and \citet{debnath2024}. 
    We also perform a local linear stability analysis, %following 
    building upon \citet{bs03}. In Section \ref{results}, we calculate growth rates of the structures in the two simulations by computing power spectra and tracking their evolution over time. In Section \ref{analysis}, we compare the growth rates from the simulations to the analytical results from the linear stability analysis and we discuss the differences between the two simulations. In Section \ref{discussion}, 
    %we discuss the differences between a Lagrangian frame and an Eulerian frame analysis and 
    we suggest possible explanations for the results we found in Section \ref{results} and we discuss possible observational implications of our results. Finally, Section \ref{conclusion} gives a summary and potential future work.
    
    \section{Background}
    \label{background}

    In this section, we outline the basic radiation-hydrodynamics (RHD) equations that are solved in this work. We also give a short overview of the key aspects of the numerical simulation methods used in \citet{moens2022} and \citet{debnath2024}, to which the reader is referred for further details. Lastly, we perform a linear stability analysis of the optically thick stellar envelope as in \citet{bs03}, bar some minor modifications for our problem set-up.
    %\JS{Ok, so here I did make q tentative suggestion for subsections, trying to make them a bit more organised, since you're using the basic equations of RHD in your stability analysis, but then you solve them differently in the actual simulations. You can modify as you wish yourself of course.}

    \subsection{Basic radiation-hydrodynamic equations} 
    %\subsection{Numerical simulation methods}

    %In this work, we analyze radiation-hydrodynamics (RHD) models produced using the RHD module of the general hydrodynamics code {\fontfamily{qcr}\selectfont MPI-AMRVAC} \citep{xia2018}. The workings of this RHD module are described in detail in \citet{moens2021}. The following set of RHD equations, 
       %JS-I replaced this for now. Feel free to do as you wish.  
    The basic equations of RHD considered in this work are
    \begin{align}
        & \partial_t \rho + \div{(\rho \vec{\varv)}} = 0, \label{eq:rhd1}\\
        & \partial_t (\rho \vec{\varv}) + \div{(\rho \vec{\varv} \vec{\varv} + p \tens{I}}) = \vec{f}_\mathrm{g} + \vec{f}_\mathrm{r}, \label{eq:rhd2}\\
        & \partial_t e + \div{(e \vec{\varv} + p \vec{\varv})} = \vec{\varv} \cdot \vec{f}_\mathrm{g} + \vec{\varv} \cdot \vec{f}_\mathrm{r} + \Dot{q}, \label{eq:rhd3}\\
        & \partial_t E + \div{( E \vec{\varv} + \vec{F}_{\mathrm{diff}}}) = - \Dot{q} - \grad{\vec{\varv}} : \tens{P}.\label{eq:rhd4}
    \end{align}
    %is solved numerically using a finite-volume method. 
    In these equations, $\vec{\varv}$ is the velocity of the gas, $\rho$ is the gas density, $p$ is the gas pressure, $\tens{P}$ is the radiation pressure tensor, $e$ is the total gas energy density, $E$ is the radiation energy density, $\vec{f}_\mathrm{g}$ is the gravitational force density, $\vec{f}_\mathrm{r}$ is the radiation force density, $\vec{F}_{\mathrm{diff}}$ is the co-moving frame (CMF) radiative flux, and $\Dot{q}$ contains contributions from heating and cooling. These equations are supplemented by the following closure relation for an ideal gas:
    \begin{equation}
        e = u + \frac{\rho \varv^2}{2} = \frac{p}{\gamma - 1} + \frac{\rho \varv^2}{2},
    \end{equation}
    where $\gamma$ is the adiabatic index and $u$ is the internal gas energy density. A constant $\gamma = 5/3$ is assumed for the gas, neglecting any ionization effects. 
    This is a reasonable assumption because the iron group ionization energy \fix{constitutes} a very small fraction of the total energy budget in these radiation-pressure dominated stars.\footnote{We note that this might change for massive stars with cooler surface temperatures for which ionization effects of hydrogen could become important for the dynamics.}
    %\fix{We make this assumption because these systems are dominated by radiation pressure and ionization effects in the gas are therefore expected to not have much effect on the dynamics of the system}. 
    The gravitational and radiation force densities are given by
    \begin{align}
        & \vec{f}_{\mathrm{g}} = - \rho \frac{G M_\star}{r^2} \hat{\vec{z}}, \\
        & \vec{f}_{\mathrm{r}} = \rho \frac{\kappa_\mathrm{F} \vec{F}_{\mathrm{diff}}}{c},
    \end{align}
    where $\kappa_\mathrm{F}$ is the flux-weighted mean opacity, $M_\star$ is the fixed stellar mass, $r$ is the distance to the point source (which corresponds to $z$ in our coordinate system), $c$ is the speed of light, and $G$ is the gravitational constant. The heating and cooling term is given by
    %\JS{Split notation in this intro-section; i.e. kappaF, kappae, kappaB} 
    \begin{equation}
        \Dot{q} = c \kappa_{\mathrm{E}} \rho E - 4 \pi \kappa_{\mathrm{P}} \rho B,
    \end{equation}
    where $B$ is the frequency-integrated Planck function, $\kappa_\mathrm{E}$ is the energy mean opacity, and $\kappa_\mathrm{P}$ is the Planck mean opacity.
    The CMF radiative flux is calculated using a flux-limited diffusion closure relation:
    \begin{equation}
        \vec{F}_{\mathrm{diff}} = - \frac{c \lambda}{\kappa_{\mathrm{F}} \rho} \grad{E}, \label{eq:fld}
    \end{equation}
    where $\lambda$ is a flux-limiter, which is used as a bridging law between the optically thick and thin analytic limits \citep{pomraning, moens2021}. This relation originates from the relation $\tens{P} = \tens{f} E$ where $\tens{f}$ is the Eddington tensor. In the optically thick limit, we may use $\tens{f} = (1/3) \tens{I}$ and $\lambda = 1/3$.
    
    \subsection{Numerical simulation methods}

    %JS-moved here for now. May reformulate to flow with text. 
    In this work, we analyze RHD models produced using the RHD module of the hydrodynamics code \textsc{mpi-amrvac} \citep{xia2018, keppens2023}. The workings of this RHD module are described in detail in \citet{moens2021}. 
    %The set of RHD equations (Eqs. \ref{eq:rhd1}-\ref{eq:rhd4}) is solved numerically using a finite volume method. 
    Following \citet{moens2022} and \citet{debnath2024}, the flux-energy and Planck mean opacities are assumed to be equal, that is, $\kappa_{\mathrm{F}} = \kappa_\mathrm{E} = \kappa_\mathrm{P} = \kappa$.
    %\JS{Above has just been moved from earlier. Here you would now first state that, following Nico and Dwaiapayan papers, you assume flux, energy, and Planck weihgthed mean opacities to be the same. Then specifically,}
    Specifically, the total opacity of the medium is calculated using a "hybrid opacity" scheme \citep{poniatowski2021} such that $\kappa = \kappa_{\text{Ross}} + \kappa_{\text{line}}$.
    Hence, the opacity is the sum of the Rosseland mean opacities $\kappa_{\text{Ross}}$ in a static medium (the OPAL opacities \citealt{opal} in this work), valid in the optically thick part of the simulations, and line opacities in a supersonic medium $\kappa_{\text{line}}$, which create a strong "line-driving" effect due to the enhancing power of Doppler shifts \citep{castor1975}. In this work, the line-driving effect plays a minor role as we focus on analyzing the optically thick sub-surface layers where $\kappa_{\text{Ross}}$ dominates.
    %\JS{I thought that was only for the linear stability analysis; in the actual simulations you have kline, right? Please reformulate (I think what you mean is that in the regions your analysis is focused kross dominates?}\\
    \\ 
    \\
    The models studied here are developed and described by \citet{moens2022} and \citet{debnath2024}, where they show simulations of WR stars and \fix{O stars}, respectively. These simulations are 2D "box-in-star" models on a grid with a 
    %radial 
    vertical direction $z$ and a lateral direction $y$. However, correction terms \fix{were} added in the vertical direction to account for sphericity effects in the divergence operator, see Appendix A in \citet{moens2021} for further details. We study representative models from both works, namely the O4 model from \citet{debnath2024} and a modified version of the $\Gamma 3$ model from \citet{moens2022}, where we extended the latter model to a lower boundary temperature of $\sim 480$ kK and applied the prescription for initial perturbations from \citet{debnath2024}. In the rest of this paper, these models \fix{are} referred to as the "O4 model" and the "WR model," respectively. The averaged fundamental parameters of both models are shown in Table \ref{table:models}. The effective temperature is defined as $F_{\mathrm{diff}}(R_{\mathrm{ph}}) \equiv \sigma T_{\mathrm{eff}}^4$, where $R_{\mathrm{ph}}$ is the photospheric radius, that is, the 
    %radial 
    vertical point where the Rosseland mean optical depth $\tau = 2/3$, and $\sigma$ is the Stefan-Boltzman constant. The luminosity is calculated as $L_\star = 4 \pi R_{\mathrm{ph}}^2 F_{\mathrm{diff}}(R_{\mathrm{ph}})$ and $L_{\rm Edd} = 4 \pi G M_\star c / \kappa_e$, with $\kappa_e = 0.34~\rm cm^2/\rm g$ for the O4 model and $\kappa_e = 0.2~\rm cm^2/\rm g$ for the (hydrogen-free) WR model. The mass loss rate is calculated as the mass flux through a spherical shell at a radial distance $z$ from the star, that is, $\Dot{M} = \langle 4 \pi {z}^2 \varv_z \rho \rangle$, with $\varv_z$ the %radial 
    vertical velocity at the outer boundary of the grid. All averages here are computed by averaging laterally and over several late time snapshots. \\
    \\
    \begin{table*}
    \caption{Fundamental parameters for the O4 model and the WR model.} \label{table:models}
    \centering 
    \begin{tabular}{c | c c c c c c c }  
    \hline\hline       
    Model & $\left<T_{\rm eff} [\mathrm{kK}]\right>$ & $M_\star/M_\odot$ & $R_{\mathrm{c}} / R_{\odot}$ & $ \langle R_{\mathrm{ph}} \rangle/R_\odot$ & $\log_{10} \left(\left<L_\star\right>/L_\odot\right)$ & $\left<L_\star\right>/L_{\rm Edd}$&  $\log_{10} \left<\Dot{M}\right>  \ [M_\odot/\mathrm{yr}] $ \\ 
    \hline                    
       $\rm O 4$ & 38.7  & 58.3 & 13.54 & 16.84 & 5.78  & 0.27 & -5.68 \\
       $\rm WR$ & 73.6  & 10.0 & 1.0 & 3.55 & 5.60  & 0.61  & -4.61\\
    \hline     
    \end{tabular}
    \tablefoot{From left to right, the columns display the model name, effective temperature, stellar mass, lower boundary radius, photospheric radius, luminosity, Eddington ratio, and mass loss rates. The angular brackets denote quantities averaged in space and time as described in the text.}
    \end{table*}    
    \begin{table}[h!]
    \caption{Grid specifications for the $\rm O 4$ model and the WR model.} \label{table:specs}      
    \centering          
    \begin{tabular}{c | c c c}  
    \hline\hline       
    Model & Cell size & Vertical domain & Lateral domain \\ 
    \hline                    
       $\rm O 4$ & $9.8 \times 10^{-4} R_{\mathrm{c}}$  & $2 R_{\mathrm{c}}$ & $0.2 R_{\mathrm{c}}$  \\
       $\rm WR$ & $4.9 \times 10^{-3} R_{\mathrm{c}}$  & $5 R_{\mathrm{c}}$ & $0.5 R_{\mathrm{c}}$ \\
    \hline                  
    \end{tabular}
    \end{table}
    A big distinguishing factor between the dynamics of the WR wind and the \fix{O star} wind is the wind launching mechanism. It has been shown that the WR star can already launch an optically thick wind from the iron bump region, where the Eddington limit is locally exceeded, and afterward line-driving takes over to further accelerate the wind \citep{poniatowski2021,moens2022}. Therefore, there is already a net outflow in the iron bump region which might impact the structure formation in this region. We discuss the effects of this mean outflow on our present study of structure formation in Appendix \ref{sec:app}. The \fix{O star}, however, launches an optically thin wind just outside the photosphere and the locally super-Eddington layer at the iron bump leads to a turbulent layer with an average velocity of zero \citep{debnath2024}. 
    Hence, the \fix{O star} is, on average, in hydrostatic equilibrium at the iron bump. A super-Eddington layer in hydrostatic equilibrium \fix{leads} to a density inversion as the radiation force exceeding gravity leads to the gas pressure gradient, and therefore the density gradient, changing sign in the hydrostatic momentum equation. Thus, the initial conditions of the O4 model contain such a density inversion. \\
    \\
    These models \fix{were} computed on a grid with four different refinement levels, but the layers we consider here lie within the highest refinement level.
    At the highest refinement level, the grid of the O4 model consists of $256 \times 2048$ cells, covering $0.2 R_{\mathrm{c}}$ laterally and $2 R_{\mathrm{c}}$ %radially. 
    vertically. The grid of the WR model consists of $128 \times 1024$ cells, covering $0.5 R_{\mathrm{c}}$ laterally and $5 R_{\mathrm{c}}$ 
    %radially. 
    vertically. Here, and for all further occurrences, $R_{\mathrm{c}}$ refers to the lower boundary radius of the respective model. The grid specifications for both models are summarized in Table \ref{table:specs} as they \fix{are} important for defining the range of perturbation wavenumbers. The cells are not perfect squares, so the cell sizes in Table \ref{table:specs} refer to their 
    %radial 
    vertical size.

    \subsection{Linear stability analysis}
    \label{lin_stability}

    The stability of the stellar envelope is studied using a linear stability analysis, %building upon the work 
    following the work by \citet{bs03}. They studied local radiation-hydrodynamic instabilities in optically thick media, which is the relevant regime for the stellar envelopes that we are studying. In this section, the key steps and results of the analysis are outlined, but the reader is referred to \citet{bs03} for further details. We also rewrite these results and inspect the physical properties of the present instabilities in the different stellar regimes that we study.\\ 
    %the full analysis. \\
    \\
    Since we \fix{study} the optically thick regime, we take $\lambda = 1/3$ in Eq. (\ref{eq:fld}) and we neglect the line driving component of the total opacity such that $\kappa = \kappa_{\rm Ross}$. We also assume the radiation pressure tensor $\tens{P}$ is isotropic and related to the radiation energy density as $\tens{P} = (1/3)E \tens{I}$. We note that \citet{bs03} did not assume all mean opacities to be equal as we do here and therefore our results are slightly different compared to them. We assume that the total state variables can be split into background values and small Eulerian perturbations (that is, perturbations at a fixed position), namely
    \begin{equation}
    \begin{split}
        &\rho \rightarrow \rho + \delta \rho, \quad \vec{\varv} \rightarrow \vec{\varv} + \delta \vec{ \varv}, \quad p \rightarrow p + \delta p, \quad \kappa \rightarrow \kappa + \delta \kappa, \\
        & T_{\mathrm{g}} \rightarrow T_{\mathrm{g}} + \delta T_{\mathrm{g}}, \quad T_{\mathrm{r}} \rightarrow T_{\mathrm{r}} + \delta T_{\mathrm{r}}.
    \end{split}
    \end{equation}
    % \begin{align}
    % & \rho \rightarrow \rho + \delta \rho, \\
    % & \vec{\varv} \rightarrow \vec{\varv} + \delta \bm { \varv}, \\
    % & p \rightarrow p + \delta p, \\
    % & \kappa \rightarrow \kappa + \delta \kappa, \\
    % & T_{\mathrm{g}} \rightarrow T_{\mathrm{g}} + \delta T_{\mathrm{g}}, \\
    % & T_{\mathrm{r}} \rightarrow T_{\mathrm{r}} + \delta T_{\mathrm{r}}.
    % \end{align}
    Here, $T_{\mathrm{g}}$ is the gas temperature, related to the gas pressure by $p = \rho k_\mathrm{B} T_\mathrm{g} / \mu$ with $k_\mathrm{B}$ the Boltzmann constant and $\mu$ the mean molecular mass, and $T_{\mathrm{r}}$ is the radiation temperature, related to the radiation energy density by $E = (4 \sigma / c) T_\mathrm{r}^4$. The background equilibrium state is assumed to be static and in local thermodynamic equilibrium (LTE). Using $\vec{\varv} = 0$ and $T_{\mathrm{g}} = T_{\mathrm{r}} = T$ in Eqs. (\ref{eq:rhd1})-(\ref{eq:rhd4}) and Eq. (\ref{eq:fld}) gives
    \begin{align}
    & - \grad{p} + \rho \vec{g} + \frac{\kappa \rho}{c}\vec{F}_{\mathrm{diff}} = 0, \label{eq:background1}\\
    & \div{\vec{F}_{\mathrm{diff}}} = 0,  \label{eq:background2}\\
    & -\frac{1}{3} \grad E - \frac{\kappa \rho}{c} \vec{F}_{\mathrm{diff}} = 0, \label{eq:background3}
    \end{align}
    as the non-trivial equations describing the equilibrium state, where we have defined the gravitational acceleration $\vec{g} = - g \hat{\vec{z}} = - (G M_\star/z^2) \hat{\vec{z}}$.
    Linearizing Eqs. (\ref{eq:rhd1})-(\ref{eq:rhd4}) and Eq. (\ref{eq:fld}) around the equilibrium state and using Eqs. (\ref{eq:background1})-(\ref{eq:background3}) to simplify gives the linearized RHD equations describing the evolution of perturbations:
    \begin{align}
    & \partial_t \delta \rho + \rho \div{\delta \vec{\varv}} + \delta \vec{\varv} \cdot \grad{\rho} = 0, \label{eq:linrhd1}\\
    & \rho \partial_t \delta \vec{\varv} = - \grad{\delta p} + \delta \rho \vec{g} - \frac{1}{3} \grad{\delta E}, \label{eq:linrhd2}\\
    & \partial_t \delta u + \delta \vec{\varv} \cdot \grad{u} + \gamma u \div{\delta \vec{\varv}} = 4 \frac{E}{T} \omega_{\mathrm{a}} (\delta T_\mathrm{r} - \delta T_{\mathrm{g}}), \label{eq:linrhd3}\\
    & \partial_t \delta E + \delta \vec{\varv} \cdot \grad{E} + \frac{4}{3} E \div{\delta \vec{\varv}} = - \div{\delta \vec{F}_{\mathrm{diff}}} - 4 \omega_{\mathrm{a}} \frac{E}{T} (\delta T_\mathrm{r} - \delta T_{\mathrm{g}}), \label{eq:linrhd4}\\
    & - \frac{1}{3} \grad{\delta E} - \frac{\kappa}{c} \delta \rho \vec{F}_{\mathrm{diff}} - \frac{\delta \kappa}{c} \rho \vec{F}_{\mathrm{diff}} - \frac{\kappa \rho}{c} \delta \vec{F}_{\mathrm{diff}} = 0 \label{eq:linrhd5},
    \end{align}
    where we have defined $\omega_{\mathrm{a}} = c \kappa \rho$ and all state variables without $\delta$ refer to the background equilibrium values. The frequency $\omega_\mathrm{a}$ describes the thermal coupling between the radiation and the gas. Due to our assumption that all mean opacities are equal, this thermal coupling frequency differs from the expression in \citet{bs03}. This is the only difference between the analysis presented here and the analysis by \citet{bs03}. Notice that Eq. (\ref{eq:linrhd3}) now describes the internal gas energy density instead of the total gas energy density. In this linearization, we assumed constant $\gamma$ and $\mu$, therefore neglecting any ionization effects and compositional gradients. \\
    \\
    We assume that the perturbations are local, which means that the perturbation wavelength should be smaller than the typical scale over which the background variables change considerably (the "short wavelength limit"). Such a typical scale is the total pressure scale height, with the total pressure being the sum of gas pressure and radiation pressure. 
    % \JS{Yes, but recall difference between gas and total pressure scale height. You have formulated the above in terms of radiation energy density E, but you may want to point out already here that in your limit for this analysis (lambad=1/3), 3P (Unity tensor)= E.}  
    We can assume a "WKB Ansatz" for such local perturbations, meaning that all perturbations have a plane wave dependence as
    \begin{equation}
        \delta \sim \exp{i (\vec{k} \cdot \vec{r} - \omega t)}.
        \label{eq:plane_wave_pert}
    \end{equation}
    Here, $\vec{k}$ is the perturbation wave vector (whose magnitude is related to the wavelength $\lambda$ as $k = 2 \pi / \lambda$) and $\omega$ is the perturbation angular frequency. In 2D, the wavevector consists of a %radial 
    vertical component $k_\mathrm{z}$ and a lateral component $k_\mathrm{y}$. Our assumption of locality means that we implicitly assume the background medium to be infinite such that no boundary specifications are needed. Generally, $\omega$ is a complex quantity, $\omega = \omega_{\mathrm{R}} + i \omega_{\mathrm{I}}$, where the real part $\omega_{\mathrm{R}}$ determines the phase speed and the imaginary part $\omega_{\mathrm{I}}$ factors out as an exponential $\sim \exp(\omega_{\mathrm{I}} t)$ which can lead to either an instability or a damped wave, with a growth or decay rate $\omega_{\mathrm{I}}$, respectively. Using Eq. (\ref{eq:plane_wave_pert}), all time derivatives and spatial derivatives of the perturbed variables can be substituted as
    \begin{equation}
        \partial_t \rightarrow - i \omega, \quad \grad \rightarrow i \vec{k}.
    \end{equation}
    Using this substitution in the linearized RHD equations, Eqs. (\ref{eq:linrhd3}) and (\ref{eq:linrhd4}) can be combined into an equation describing the perturbation in total pressure, Eq. (40) in \citet{bs03}. We consider the case in which the radiation and the gas are tightly thermally coupled, that is, the limit $\omega_\mathrm{a} \rightarrow \infty$. This limit is applicable here since a typical coupling time $1/\omega_{\mathrm{a}}$ in the relevant regions of the studied models is one to two orders of magnitude smaller than the photon free-flight time across the lateral extent of the computational domain, which means that the radiation and the gas are indeed tightly coupled. Further taking the limit $k \rightarrow \infty$ as we are considering local perturbations, this equation can then be rewritten and factorized into a dispersion relation (which is equivalent to Eq. (59) in \citet{bs03})
    \begin{multline}
        \label{eq:factorization}
         \left( \omega + \frac{i c k^2}{3 \kappa \rho} \frac{4 ( \gamma -1 ) E}{p + 4 ( \gamma -1) E} \right) \left(\omega^2 - k^2 c_{\mathrm{i}}^2\right) \\
        \left[\omega + i \frac{3 \kappa \rho}{c k^2} \left(1 - \left(\hat{\vec{k}} \cdot \hat{\vec{z}}\right)^2\right) \left( 1 + \frac{4 E}{3 p }\right) \left(\frac{\gamma p}{4(\gamma - 1)E} N_{\mathrm{g}}^2 + \frac{1}{3} N_{\mathrm{r}}^2 \right) \right] = 0,
    \end{multline}
    where we only retain the highest order terms in $k$ for each mode here, $c_{\mathrm{i}} = \sqrt{ p/\rho }$ is the isothermal gas sound speed, $\hat{\vec{k}} = \vec{k}/k$ and
    \begin{align}
        & N_{\mathrm{g}}^2 = -\vec{g} \cdot \big( \frac{1}{\rho c_{\mathrm{g}}^2} \grad{p} - \grad{\ln \rho} \big) =  -\frac{(\gamma - 1) \rho T}{\gamma p} \vec{g} \cdot \grad{S_{\mathrm{g}}}, \label{eq:brunt_gas}\\
        & N_{\mathrm{r}}^2 = -\vec{g} \cdot \big( \frac{1}{3\rho c_{\mathrm{r}}^2} \grad{E} - \grad{\ln \rho} \big) = -\frac{3 \rho T}{4 E} \vec{g} \cdot \grad{S_{\mathrm{r}}}, \label{eq:brunt_rad}
    \end{align}
    are the Brunt-V\"ais\"al\"a frequencies in gas and radiation, respectively, with the adiabatic gas sound speed $c_{\mathrm{g}} = \sqrt{\gamma p/\rho }$ and the radiation sound speed $c_{\mathrm{r}} = \sqrt{4 E/(9\rho) }$, and 
    \begin{align}
        & S_{\mathrm{g}} = \frac{k_\mathrm{B}}{\mu (\gamma -1)} \ln{(p \rho^{-\gamma})} + \mathrm{constant}, \\
        & S_{\mathrm{r}} = \frac{4 E}{3 \rho T} + \mathrm{constant},
    \end{align}
    are the gas and radiation entropy.
    This dispersion relation describes four possible wave modes, out of which three modes can potentially become unstable. Albeit, two unstable modes are only recovered when including lower order corrections in k (see Eq. \ref{eq:freq_ac}). We make a distinction in terminology between %propagating and non-propagating modes.
    modes with $\Re(\omega) \neq 0$ or $\Re(\omega) = 0$.
    A %non-propagating 
    mode with $\Re(\omega) = 0$ and increasing or decreasing amplitude \fix{is} referred to as "unstable" or "damped," respectively, while a %propagating 
    mode with $\Re(\omega) \neq 0$ and increasing or decreasing amplitude \fix{is} referred to as "overstable" or "overdamped," respectively.
    %The first factor in Eq. (\ref{eq:factorization}) describes upwards and downwards propagating acoustic waves in the radiating fluid. 
    The first factor in Eq. (\ref{eq:factorization}) describes a radiative diffusion mode that is purely damped (and is therefore not of interest in the current study of structure growth) because of the negative, purely imaginary frequency. The last factor in Eq. (\ref{eq:factorization}) describes non-propagating gravity waves with frequency
    \begin{equation}
        \label{eq:freq_grav}
        \omega_{\mathrm{g}} = - i \frac{3 \kappa \rho}{c k^2} \left(1 - \left(\hat{\vec{k}} \cdot \hat{\vec{z}}\right)^2\right) \left( 1 + \frac{4 E}{3 p }\right) \left(\frac{\gamma p}{4(\gamma - 1)E} N_{\mathrm{g}}^2 + \frac{1}{3} N_{\mathrm{r}}^2 \right).
    \end{equation} 
    Using Eqs. (\ref{eq:brunt_gas}) and (\ref{eq:brunt_rad}), we can rewrite 
    \begin{equation}
        \left( 1 + \frac{4 E}{3 p }\right) \left(\frac{\gamma p}{4(\gamma - 1)E} N_{\mathrm{g}}^2 + \frac{1}{3} N_{\mathrm{r}}^2 \right) = -\left( \frac{\gamma T}{3 c_{\mathrm{g}}^2} + \frac{T}{9 c_{\mathrm{r}}^2} \right) \vec{g} \cdot \grad{S_\mathrm{tot}},
    \end{equation}
    with $S_{\mathrm{tot}} = S_{\mathrm{g}} + S_{\mathrm{r}}$. This product can therefore be identified with the total Brunt-V\"ais\"al\"a frequency $N_{\mathrm{tot}}^2$. Furthermore, we also recognize the first factor in Eq. (\ref{eq:freq_grav}) as the radiative diffusion coefficient $D = c/(3 \kappa \rho)$. Thus, the total gravity wave frequency can be rewritten as
    \begin{equation}
        \omega_{\mathrm{g}} = -\frac{i}{D k^2} \left(1 - \left(\hat{\vec{k}} \cdot \hat{\vec{z}}\right)^2\right) N_{\mathrm{tot}}^2.
    \end{equation}
    The factor $1/(D k^2)$ defines a radiation diffusion time $t_{\mathrm{diff}}$ over the perturbation wavelength $\lambda = 2\pi / k$ and the total Brunt-V\"ais\"al\"a frequency defines the growth or decay time $t_{\mathrm{g}}^2 =1/N_{\mathrm{tot}}^2$ the perturbation would have when neglecting the effects of radiative diffusion. We note that $t_{\mathrm{g}}^2 < 0$ for growth and $t_{\mathrm{g}}^2 > 0$ for decay. So,
    \begin{equation}
        \omega_{\mathrm{g}} \propto -\frac{t_{\mathrm{diff}}}{t_{\mathrm{g}}^2}.
        \label{eq:con_timescales}
    \end{equation}
    From this scaling, it is clear that the growth rate of the gravity mode \fix{decreases} with decreasing $t_{\mathrm{diff}}$, namely the perturbations are quickly washed away by rapid radiative diffusion. This diffusion time is itself dependent on the perturbation wavenumber which gives the final $\omega_{\mathrm{g}} \sim 1/k^2$ dependence. This dependence can also be seen by rewriting the diffusion timescale as $t_{\mathrm{diff}} \sim t_{\mathrm{ff}} \tau_\lambda$, where $t_{\mathrm{ff}} = \lambda/c$ is the photon free-flight timescale over the perturbation wavelength $\lambda$ and $\tau_\lambda = \kappa \rho \lambda$ is the associated optical depth for this length. As expected, radiative diffusion enhances the photon escape time with a factor $\sim \tau_\lambda$.
    %\JS{Suggest to expand this physically interesting explanation a bit, perhaps something along lines: 'This dependence can also be seen by re-writing the diffusion time-scale as $t_{\mathrm{diff}} \sim t_{\mathrm{ff}} \tau_\lambda$, where $t_{\mathrm{ff}} = \lambda/c$ is the photon free-flight time-scale over the perturbation wavlength $\lambda$ and $\tau_\lambda = \kappa \rho \lambda$ the associated optical depth for this length. That is, as expected radiative diffusion enhances the photon escape time with a factor $\sim \tau$.} 
    Perturbations with shorter wavelengths (and larger wavenumbers) thus experience quicker radiative diffusion and therefore have smaller growth rates. 
    The gravity mode has a purely imaginary frequency, which means that this mode \fix{either} decays ($\omega_{\mathrm{I}} < 0$) or grows exponentially ($\omega_{\mathrm{I}} > 0$). The frequency of the gravity mode can have a changing sign depending on the gradients of the background state variables. Therefore, the gravity modes \fix{are} unstable if $N_{\mathrm{tot}}^2 < 0$, or equivalently, 
    \begin{equation}
        \label{eq:stab_con}
        \frac{\gamma p}{4(\gamma - 1)E} N_{\mathrm{g}}^2 + \frac{1}{3} N_{\mathrm{r}}^2 < 0.
    \end{equation}
    %\JS{Finally, if possible, it'd be nice to see how this corresponds to classical Schwarchsild criterion here, before moving to acoustic instability.} 
    Finally, the second factor in Eq. (\ref{eq:factorization}) describes, to highest order in $k$, two isothermal acoustic wave modes traveling in opposite directions. When including first order corrections to the frequencies, it can be found that these isothermal acoustic modes have a non-zero $\omega_{\mathrm{I}}$ as well (see also \citealt{bs03}, their Eq. 62), namely
    \begin{equation}
        \label{eq:freq_ac}
        \omega_{\mathrm{ac}} = \pm k c_{\mathrm{i}} - i \frac{\kappa}{2 c c_{\mathrm{i}}} \bigg( 1 + \frac{3 p }{4 E} \bigg) \left[ \left( \frac{4 E}{3} + p \right)c_{\mathrm{i}} \mp \left(\hat{\vec{k}} \cdot \vec{F}_{\mathrm{diff}} \right)\Theta_{\rho}\right],
    \end{equation}
    where the upper and lower signs describe upward and downward propagating waves, respectively, and $\Theta_{\rho} = \partial \ln \kappa/\partial \ln \rho$. The instability criterion for these modes is
    \begin{equation}
        \label{eq:stab_ac}
        \left( \frac{4 E}{3} + p \right)c_{\mathrm{i}} \mp \left(\hat{\vec{k}} \cdot \vec{F}_{\mathrm{diff}} \right)\Theta_{\rho} < 0.
    \end{equation}
    The stability of the different propagation directions depends on the sign of $\Theta_{\rho}$. For $\Theta_{\rho} > 0$, the upward propagating waves can become overstable while the downward propagating waves are overdamped. These overstable acoustic modes are believed to be a local version of the global so-called strange mode instability \citep{gautschyandglatzel,glatzel1994,saio} and \fix{we therefore refer to this instability} as "strange mode" from now on. It is important to note that these strange modes can only become overstable when there are non-zero opacity derivatives and therefore, there would be no instability in a pure Thomson scattering medium. This is in apparent contradiction with \citet{shaviv2001} who found these instabilities in purely Thomson atmospheres, although it is important to note that he performed a global analysis as opposed to this local analysis.\\
    \\   
     \begin{figure*}[h!]
        \centering
        \includegraphics[width=18cm]{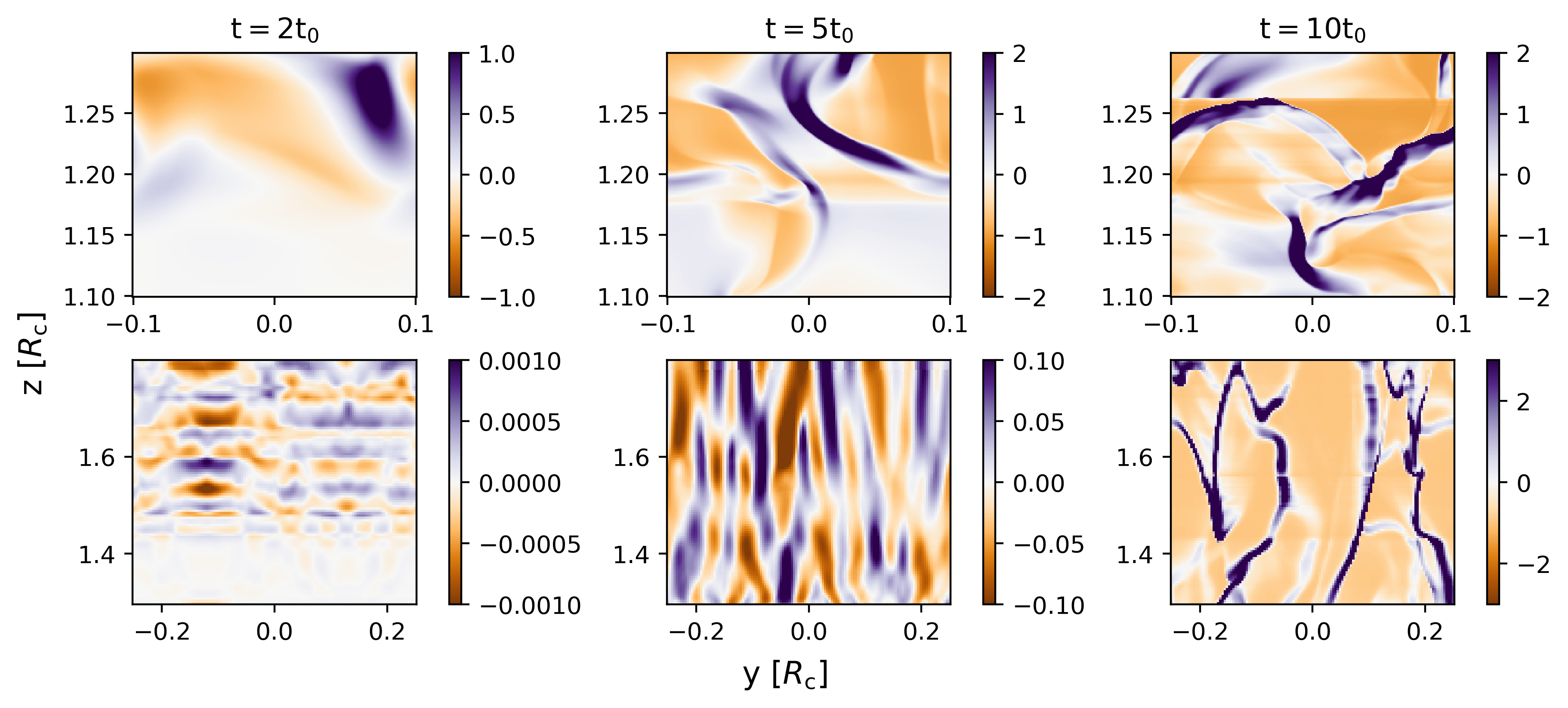}
        \caption{Relative density $\Delta_{\rho}$ (as defined in the text) at $t= 2 t_0$, $t= 5 t_0$ and $t= 10 t_0$ for the O4 model (top) and the WR model (bottom). Here, $t_0$ refers to the models' respective dynamical timescales of $t_0 \approx 5275$ s for the O4 model and $t_0 \approx 824$ s for the WR model. The colorbars do not necessarily have the same range.}
        \label{fig:rel_dens}
    \end{figure*}
    Following \citet{Owocki_2014}, we can rewrite Eq. (\ref{eq:freq_ac}) by regrouping some terms into a "radiation drag" term ${D}_\mathrm{r}$ such that the imaginary part of $\omega_\mathrm{ac}$ becomes
    \begin{equation}
        \Im(\omega_{\mathrm{ac}}) = \frac{g_{\mathrm{r}}}{2 c_{\mathrm{i}}} \left(1 + \frac{3 p}{4 E}\right) \left( \pm \Theta_{\rho} - {D}_\mathrm{r} \right),
    \end{equation}
    with 
    \begin{equation}
        {D}_\mathrm{r} = \left(\frac{4 E}{3} + p\right) \frac{c_\mathrm{i}}{F_\mathrm{diff}}.
    \end{equation}
    and $g_\mathrm{r} = \kappa F_\mathrm{diff}/c$.
     Using $g_{\mathrm{r}} = \Gamma g$, with $\Gamma$ the Eddington factor, and taking the second factor to be close to one in radiation-pressure dominated media, we can further rewrite
    \begin{equation}
        \Im(\omega_{\mathrm{ac}}) \approx \frac{\Gamma g}{2 c_{\mathrm{i}}} \left( \pm \Theta_{\rho} - D_\mathrm{r} \right).
        \label{eq:ac_drag}
    \end{equation}
    The first factor defines a certain growth time $t_\mathrm{ac}$, ${\Gamma g}/(2 c_{\mathrm{i}}) \sim 1 / t_\mathrm{ac}$, which is a function of $\Gamma$. The total growth time of the acoustic instability depends on $t_\mathrm{ac}$ and on the competition between $\Theta_{\rho}$ (the "driving term") and $D_\mathrm{r}$ (the "damping term"). Thus, in order to have an instability, the driving term $\abs{\Theta_{\rho}}$ must be larger than the damping term $D_\mathrm{r}$, which is equivalent to the criterion of Eq. (\ref{eq:stab_ac}). This $\Theta_{\rho}$ depends on the opacity and is therefore also a function of metallicity and helium abundance. The radiation drag $D_\mathrm{r}$ depends on the flux $F_\mathrm{diff}$ and hence, assuming a gas dominated by radiation pressure, $D_\mathrm{r} \sim E c_{\mathrm{i}} /F_\mathrm{diff} \sim (c_{\mathrm{i}}/c) (T/\Tilde{T}_{\mathrm{eff}})^{4} $, where we have defined a local %`effective' 
    flux temperature $\sigma \Tilde{T}_{\mathrm{eff}}^{4} \equiv F_\mathrm{diff}(z_0) \sim L_\star/z_0^2$ at the iron bump (located at $z_0$). Since this %effective 
    flux temperature is much higher for WR stars than for \fix{O stars}, whereas $T$, $c_{\mathrm{i}}/c$ and $\Theta_{\rho}$ are similar for the considered region, these scalings %directly 
    suggest that this instability should be more effective in WR stars than in \fix{O stars}. 
    %\JS{NOTE: this is actually very similar to the scaling relations I did for evaluating 'efficiency' of convection in the \fix{O star} paper Dwaipayan led, see Section 5.1. Not sure what this means actually. Addition: It's quite interesting to note that (under assumption that radiatvie enthalpy dominates potential 'convective' energy transport) one could write the scaling of Dtilde as Dtilde propto Fconv/Fdiff. That is, whenever enthalpy ('convection') is effecient in transportint *energy*, this instability will be heavily damped, or vanish. But the less efficient energy transport by enthalpy is in comparison to radiative diffusion, the less of a role this damping / drag will be. This actually does explain why it is less important in WR stars than O stars.} 
    \\
    \\
    From this rewriting, we can also see that for lower luminosity stars, with lower $\Gamma$, $t_\mathrm{ac}$ \fix{is} larger and therefore the total growth time \fix{is also} larger than for higher luminosity stars. This means that this instability is less relevant in the lower luminosity regime. It is also important to reiterate that the strange modes are propagating waves as opposed to the gravity modes, which means that any structure growth initiated by this mode \fix{does} not necessarily manifest as structures at a fixed point in space. Finally, we see that the imaginary part of Eq. (\ref{eq:freq_ac}) is independent of wavenumber, that is, $\Im(\omega_\mathrm{a}) \sim k^0$, which is a different scaling than what was found for the gravity mode. \\
    \\
    Due to the discrete nature of the grid-based simulations studied here, the wavenumbers $k$ are a discrete set of values. The upper and lower limits are set by the cell size $L_{\mathrm{cell}}$ and the box size $L_{\mathrm{box}}$, respectively, namely $k \geq 2\pi / L_{\text{box}} $ and $k \leq \pi / L_{\text{cell}} $. However, since the analysis above is only valid in the short wavelength limit, we must further restrict the range of wavenumbers to be analyzed. As mentioned earlier, the WKB Ansatz is valid only when background variables do not vary considerably over the perturbation wavelength. We \fix{took} the total pressure scale height $H = \beta c_{\mathrm{g}}^2 / g$, with $\beta = (p + E/3) / p$, as a typical length scale over which the background values vary. The iron bump region is the relevant region for instabilities to develop, so we \fix{calculated} a characteristic scale height $H_0$ of a single snapshot as the scale height at the %radial 
    vertical point where the OPAL opacity reaches its maximum. 
    %\JS{Is this located from initial conditions, on a snapshot, or on a time-average?} T
    Therefore, the lower limit for the wavenumbers becomes $k \geq \max[2 \pi / L_{\mathrm{box}}, 2 \pi / H_0$]. We also \fix{defined} a characteristic dynamical timescale as $t_0 = H_0 / c_{\mathrm{r},0}$, where the radiation sound speed \fix{was} evaluated at the same point as the scale height. We \fix{used} the radiation sound speed to define the dynamical timescale because the system is radiation-dominated and hence the radiation sound speed is larger than the gas sound speed. %\JS{I did not see definition of radiatioon sound speed above, or did I miss it? Also, you need to motivate use of that instead of gas sound speed (radiation dominated atmospheres, E/3 gt p).
    
    \section{Results}
    \label{results}
    %\subsection{Single perturbation simulations}

\begin{figure}[t!]
        \centering
        \includegraphics[width=9cm]{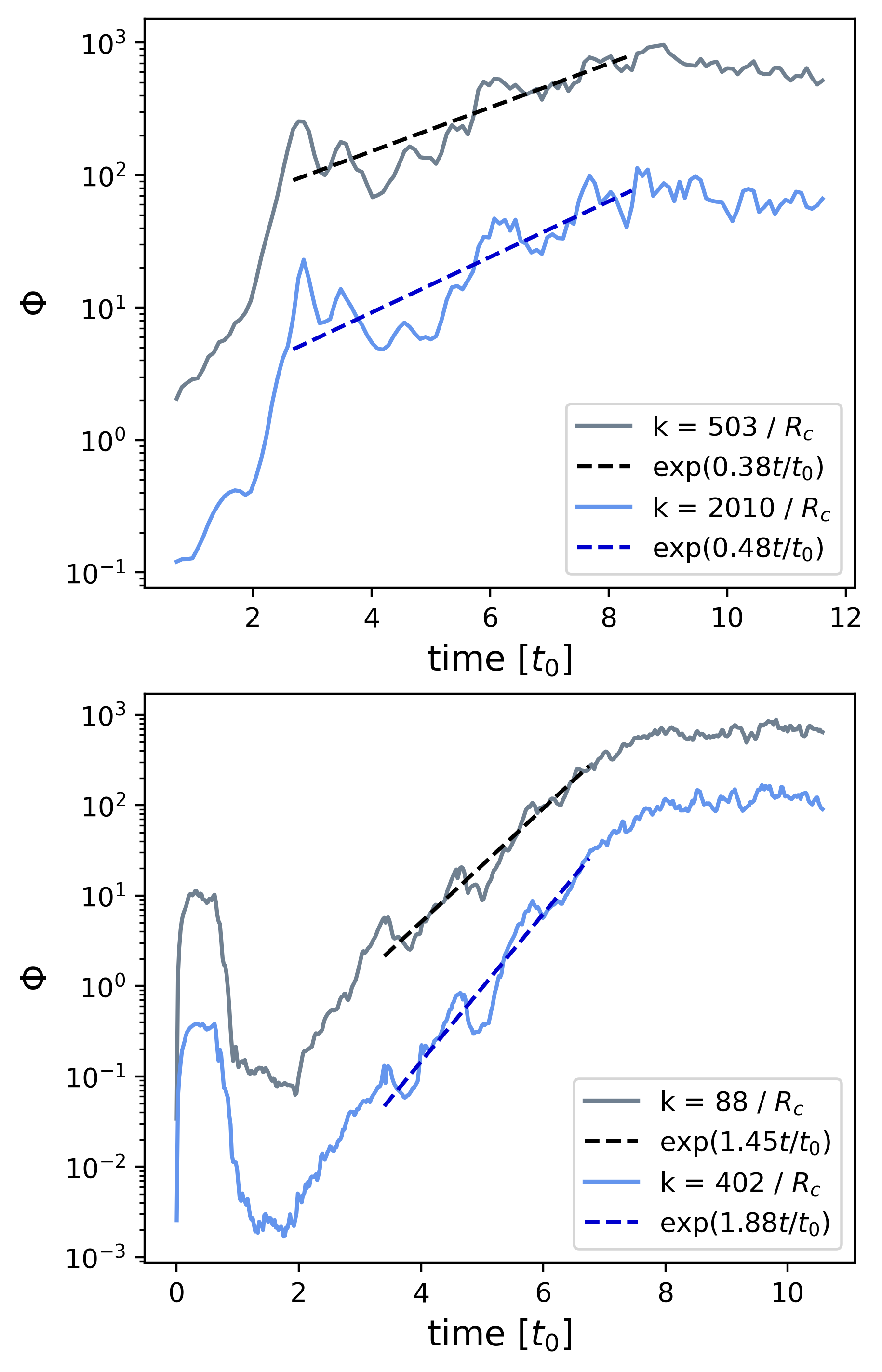}
        \caption{Density power evolution over time of the different wavenumber modes in the O4 model (top) and the WR model (bottom). The blue curves show the evolution of the large-wavenumber mode and the gray curves show the small-wavenumber mode. The dashed curves show the respective exponential fits.}
        \label{fig:density_power}
    \end{figure}

    %\subsection{Density power spectra}
    We \fix{computed} power spectra to quantify the evolution of structures in the presented simulations.
    Defining the relative density as 
    \begin{equation}
        \Delta_{\rho}(\vec{r}) = \frac{\rho - \langle \rho \rangle}{\langle \rho \rangle},
    \end{equation}
    where $ \langle \rho \rangle $ is the laterally averaged density at a radial point, the density power spectrum is given by
    \begin{equation}
        \Phi(k) = \left\langle\sqrt{ \hat{\Delta}_{\rho}^{*}(\vec{k}) \hat{\Delta}_{\rho}(\vec{k})}\right\rangle,
    \end{equation}
    where $\hat{\Delta}_{\rho}(\vec{k})$ is the 2D Fourier transform of the relative density and the averaging is done over circles of equal $k$. Previously, the plane-wave dependence of the perturbations was already introduced in Eq. (\ref{eq:plane_wave_pert}). However, a more complete treatment would be to expand a perturbation in Fourier modes as
    \begin{equation}
        \delta = \sum_{\vec{k}} \hat{\delta}(\vec{k}) \exp{ i (\vec{k} \cdot \vec{r} - \omega_{\mathrm{R}} t)},
    \end{equation}
    where the imaginary part of the frequency $\omega_{\mathrm{I}}$ has been absorbed in the Fourier coefficients $\hat{\delta}(\vec{k})$. Therefore, the time evolution of the power spectrum $\Phi(k)$ for any particular wavenumber is expected to follow an exponential curve with a certain growth rate. These growth rates can then be compared to analytic $\omega_{\mathrm{I}}$ for the convective mode and the strange mode to distinguish between these two instabilities. First, we \fix{calculated} the characteristic scale height and dynamical timescale, as defined above, at the time when the perturbations start experiencing steady exponential growth. For the WR model, we get $H_0 = 0.80$$R_{\mathrm{c}}$ and $t_0 \approx 824$ s. Therefore, we get approximately $k \in \left[ 13 / R_{\mathrm{c}}, 641 / R_{\mathrm{c}} \right] $ as the range of resolvable wavenumbers in the WR simulation. 
    %\JS{One might criticise the use of pressure scale height for WR star, since it involves an outflow in the considered region. Is a density/velocity scale height rho/drho/dz approx v/dv/dz of the same order?} \CV{From a quick calculation, the density scale height seems to be about a factor of 5 smaller.} 
    For the O4 model, we get $H_0 = 0.13 R_{\mathrm{c}}$ and $t_0 \approx 5275 s$. Therefore, we get approximately $k \in  \left[49 / R_{\mathrm{c}}, 3204 / R_{\mathrm{c}}\right] $ for the wavenumber range. 
    In Fig. \ref{fig:rel_dens}, we show the relative density as defined above for three different time snapshots in the O4 model (top row) and the WR model (bottom row). In this figure, one can indeed nicely see the development of structure. \\
    \\
    % \begin{figure}[t!]
    %     \centering
    %     \includegraphics[width=9cm]{O_G3_time_evolution.png}
    %     \caption{Density power evolution over time of the mode with wavelength equal to the scale height (grey curve) and the mode with wavelength equal to a tenth of the scale height (blue curve) in the O4 model. The dashed lines show exponentials with respective best fit growth rates.}
    %     \label{fig:O_time}
    % \end{figure}
    % \begin{figure}[t!]
    %     \centering
    %     \includegraphics[width=9cm]{WR_G5_time_evolution.png}
    %     \caption{Density power evolution over time of the mode with wavelength equal to the scale height (grey curve) and the mode with wavelength equal to a tenth of the scale height (blue curve) in the WR model. The dashed lines show exponentials with respective best fit growth rates.}
    %     \label{fig:WR_time}
    % \end{figure}
    In the top panel of Fig. \ref{fig:density_power}, the evolution of the density power $\Phi$ for two different wavenumbers, namely $k = 503 / R_\mathrm{c}$ and $k = 2010 / R_\mathrm{c}$, in the O4 simulation is shown. To avoid creating a bias in a certain direction, we cut the original grid into a square from $z = 1.1 R_{\mathrm{c}}$ to $z = 1.3 R_{\mathrm{c}}$ and $y = -0.1 R_{\mathrm{c}}$ to $y = 0.1 R_{\mathrm{c}}$. The %radial 
    vertical range was chosen to accurately resolve the iron opacity peak and to lie well within the convectively unstable region, see further discussion in Section \ref{analysis_ostar}. The growth rates for these two modes can be deduced by fitting an exponential $\sim \exp(\omega t)$ to these curves. The fitting \fix{was} done starting from the time snapshot where steady exponential growth starts ($t \approx 2 t_0$) and until a plateau is reached ($t \approx 8 t_0$). We find growth rates $0.38 / t_0$ and $0.48 / t_0$ for the small-wavenumber mode and the large-wavenumber mode, respectively. The result for the WR model is shown in the bottom panel of Fig. \ref{fig:density_power}. As before, we cut the grid into a square, now from $z = 1.3 R_{\mathrm{c}}$ to $z = 1.8 R_{\mathrm{c}}$. Both modes seem to be damped initially, until around $t \sim 2 t_0$. This is likely due to the perturbations in the initial conditions being washed away by the presence of an outflow in this region. Afterward, the growth continues steadily with growth rates $1.45 / t_0$ and $1.88 / t_0$ for the small-wavenumber mode with $k = 88 / R_\mathrm{c}$ and the large-wavenumber mode with $k = 402 / R_\mathrm{c}$, respectively. Importantly, the growth reaches a plateau in both simulations, indicating that a full turbulent state is reached. This plateau seems to be at higher value for the small wavenumber modes in both cases. \\
    \\
    In \citet{jiang2015}, a similar initial damping as in our WR model is found for the large-wavenumber modes in their StarTop model (which has similar stellar parameters as our O4 model). Afterward, they see quick exponential growth which they attribute to a non-linear cascade. However, as the initial damping in the WR model seems to be present irrespective of wavenumber and we do not see the same behavior in the O4 model, we deem such a non-linear cascade an unlikely explanation for the initial structure growth in our models. \\
    \\

    \begin{figure}[t!]
        \centering
        \includegraphics[width=9cm]{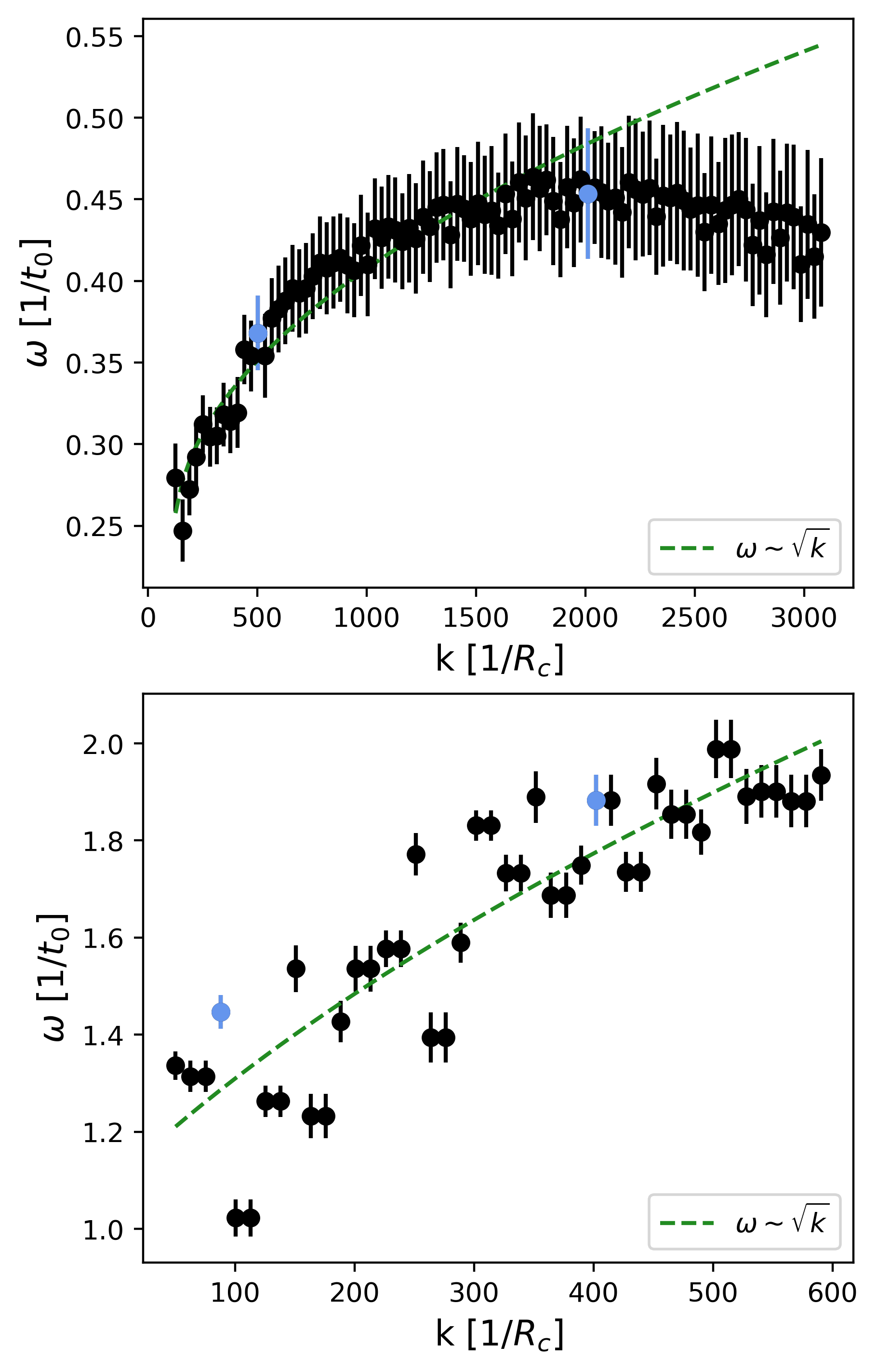}
        \caption{Growth rates calculated from the exponential fit of the density power evolution for multiple wavenumbers in the O4 model (top) and WR model (bottom). The blue dots correspond to the wavenumbers shown in Fig. \ref{fig:density_power}. The errorbars on the growth rates are one standard deviation errors from the nonlinear least squares fitting. The green dashed line is a fit of $\omega(k)$.}
        \label{fig:rates_vs_k}
    \end{figure}
    As discussed in Section \ref{lin_stability}, the growth rates for both the convective instability and the strange mode instability have a particular $k$-dependence. The growth rates, computed from the power evolution curves, can be calculated for a range of wavenumbers in order to compare to theory.
    In Fig. \ref{fig:rates_vs_k}, the resulting growth rates from the O4 model (top panel) and the WR model (bottom panel) are shown as a function of wavenumber. We omitted the first and last three wavenumbers from their respective wavenumber range to avoid any resolution effects. The points in blue indicate the rates found from the curves in Fig. \ref{fig:density_power}. %It is clear that the results of the O4 model do not exhibit the $\sim 1/k^2$ dependence of the convective instability nor are they independent of wavenumber as would be the case for the strange mode. 
    It is clear that neither the O4 model nor the WR model shows the expected $1/k^2$ dependence of the convective instability. \\
    \\
    In fact, in the O4 model, up to $k \approx 2000 / R_\mathrm{c}$, the growth rates increase with increasing wavenumbers, approximately as $\sim \sqrt{k}$. A fit of $\omega \sim \sqrt{k}$ up to $k = 2000 / R_\mathrm{c}$ is shown in the figure as the green dashed line. Afterward, there seems to be a plateau or a lightly decreasing trend and a clear deviation from this $\omega \sim \sqrt{k}$ behavior. This decrease for higher wavenumbers could be related to numerical damping being largest at smaller scales. %\JS{I guess this part below is in need of a small update now?} 
    % In the WR model, however, there does not seem to be a clear trend, as can be seen in Fig. \ref{fig:rates_vs_k}. Instead, the values seem to fluctuate around a mean value $\sim 1.33 / t_0$, indicated by the red dashed line. Furthermore, the growth rates from the WR model are, overall, larger than the ones from the O4 model as well. The different dependence on wavenumber and the different magnitudes of $\omega$ suggest that the mechanisms for creating these structures are different in the WR model as compared to the O4 model. \JS{Could reformulate to emphasize more the key result that neither case show any sign of characteristic 1/k-sq dependence predicted from the convective instability.} Furthermore, the $k$-dependence from the O4 model does not point to either of the instabilities studied here (see further discussion in section \ref{analysis}) while the $k$-dependence of the WR model could be consistent with the strange mode instability.
    In the WR model, this same $\omega \sim \sqrt{k}$ behavior is found, although the spread in growth rates is larger than for the O4 model. We also find no deviation from this trend at higher wavenumbers. It has to be noted that, due to the larger spread, the $k$-dependence is less clear as a linear dependence, for example, could in principle fit the data as well. In any case, the WR model also shows no clear agreement with either instability discussed in Section \ref{lin_stability}. Interestingly, the growth rates in the WR model are overall larger than the growth rates in the O4 model. To really see the difference between the growth rates from the different models, we need to translate them to the same units. Comparing the mode with $k = 88 / R_\mathrm{c}$ in the WR model to the mode with $k = 2010 / R_\mathrm{c}$ in the O4 model, as these are approximately the same wavenumber if we translate them to the same units, we find a growth rate $\omega = 1.8 \times 10^{-3}$ $\mathrm{s}^{-1}$ in the WR model and $\omega = 8.5 \times 10^{-5}$ $\mathrm{s}^{-1}$ in the O4 model. Thus, for similar wavenumber modes, the growth rate in the WR model is around a factor of twenty larger than the growth rate in the O4 model. This could also explain why there is no drop-off at higher wavenumbers as small numerical damping would have less of an effect on the large growth rates of the WR model. %Importantly, neither model shows the $1/k^2$ behavior expected for the convective instability which, as mentioned before, is classically believed to be the driver of turbulence in these envelopes.

    % \begin{figure}[t!]
    %     \centering
    %     \includegraphics[width=9cm]{O_om_vs_k.png}
    %     \caption{Growth rate of the exponential fit of the time evolution of density power for multiple wavenumbers in the O4 model. \JS{So from a (rather quick) look, it seems basic dependency of growth of RT instability should indeed be with increasing wavenumber... specifically it seems to go as w propto sqrt(k). You think that's roughly consistent with this growth?} \color{red} CV: add explanation on error bar calculation}
    %     \label{fig:O_om_vs_k}
    % \end{figure}
    %  \begin{figure}[t!]
    %     \centering
    %     \includegraphics[width=9cm]{WR_om_vs_k.png}
    %     \caption{Growth rate of the exponential fit of the time evolution of density power for multiple wavenumbers in the WR model. The red dashed line shows the mean value. \color{red} CV: add explanation on error bar calculation}
    %     \label{fig:WR_om_vs_k}
    % \end{figure}

    \section{Analysis}
    \label{analysis}
    
    Using Eq. (\ref{eq:stab_con}) and Eq. (\ref{eq:stab_ac}), we can predict the unstable regions for the convective mode and the strange mode for both models. Moreover, the imaginary parts $\omega_{\mathrm{I}}$  of the resulting frequencies in Eqs. (\ref{eq:freq_grav}) and (\ref{eq:freq_ac}) for these modes allow us to estimate the growth rates set by both instabilities. Eqs. (\ref{eq:freq_grav}) and (\ref{eq:freq_ac}) depend on the direction of the wave vector so whenever we calculate characteristic values for the growth rates, we assume that the wave vector makes a $45^\circ$ angle with the 
    %radial 
    vertical direction.

    \subsection{\fix{O stars}}
    \label{analysis_ostar}

\begin{figure}
        \centering
        \includegraphics[width=9cm]{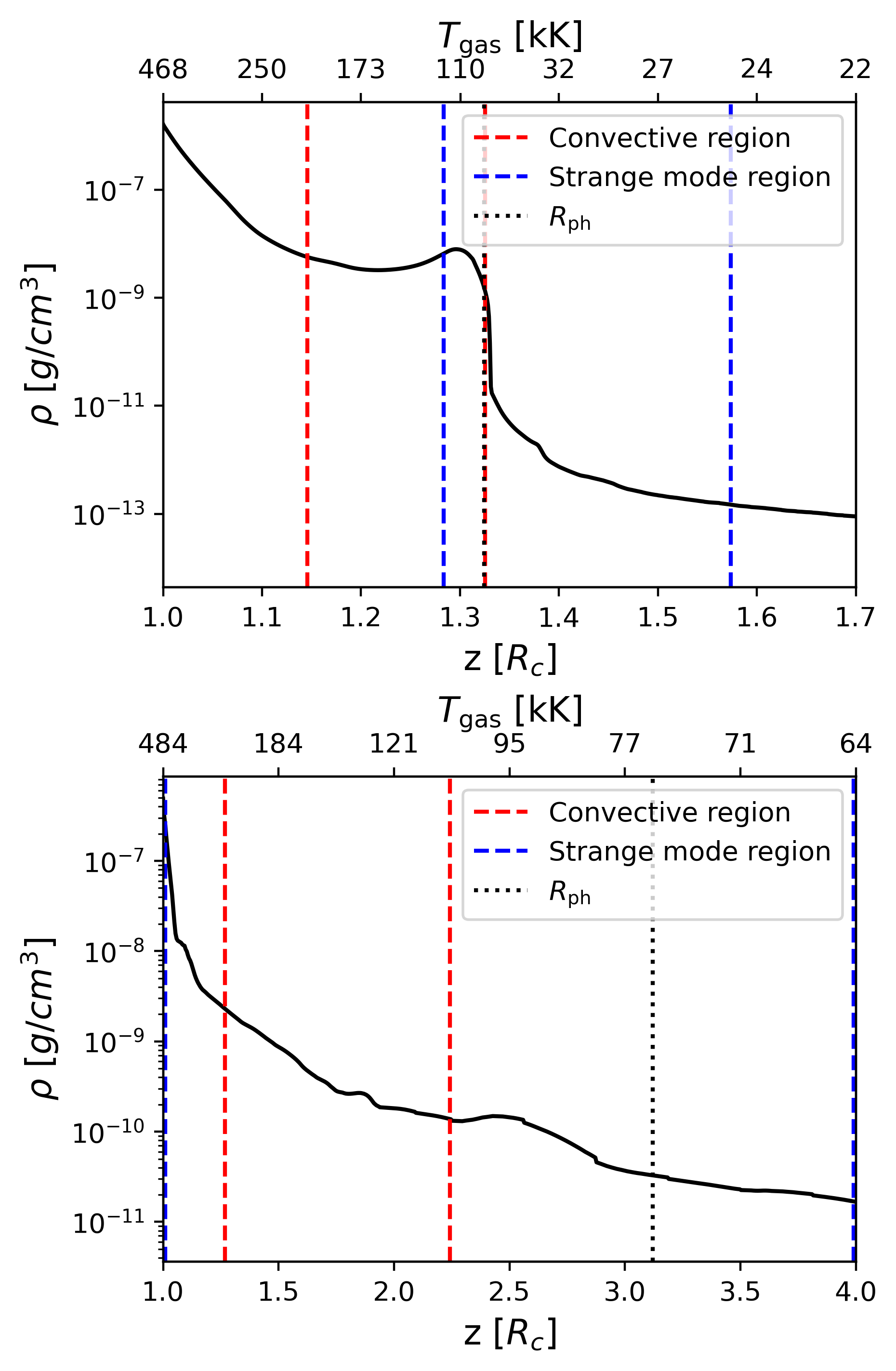}
        \caption{Average density profiles at $t=2 t_0$ for the O4 model (top) and the WR model (bottom), with the vertical axis in logarithmic scale. The red dashed lines indicate the predicted convective region and the blue dashed lines indicate the predicted strange mode region. In the WR model, the whole domain is predicted to be unstable to strange modes, which we indicate by the blue dashed lines at the edges of the plot. The dotted black lines indicate the photosphere.}
        \label{fig:init}
    \end{figure}

    In the top panel of Fig. \ref{fig:init}, we show the laterally averaged density structure for the snapshot at $t = 2 t_0$ of the O4 model, which corresponds to the time where we started the exponential fit in the top panel of Fig. \ref{fig:density_power}. With the red and blue dashed lines, we show the edges of the regions where the convective mode and the strange mode become unstable or overstable, respectively. We note that there is some overlap between the convective region and the strange mode region. 
    %\JS{This now requires that you have made connection to strange modes before; I think you have only described them as acoustic modes until now.} 
    However, the convective region is located completely below the photosphere while the strange mode region starts closer to the photosphere and extends into the wind. It must be noted that these regions are predicted using analytic results that are derived in the optically thick limit which means that these criteria can really only be applied in the sub-surface layers. Therefore, we focus solely on the sub-surface region, dominated by the convective instability in the O4 model, albeit with a smaller strange mode region also. The growth rates we calculate from Eqs. (\ref{eq:freq_grav}) and (\ref{eq:freq_ac}) depend on the local background variables at every grid point, but as a characteristic value, we can take the mean value over all grid points in the same region where we computed the power evolution (Fig. \ref{fig:density_power}) in the previous section, namely from $z=1.1 R_{\mathrm{c}}$ to $z=1.3 R_{\mathrm{c}}$ and $y = -0.1 R_\mathrm{c}$ to $y = 0.1 R_\mathrm{c}$. For $k \approx 503 / R_\mathrm{c}$, the small wavenumber mode in the top panel of Fig. \ref{fig:density_power}, this results in a mean convective growth rate of $\langle\omega_{\mathrm{g}}\rangle \approx 0.0014 / t_0$. The growth rate for the strange mode can also be computed, calculating the opacity gradient in Eq. (\ref{eq:freq_ac}) directly from the opacity table used in the model. The growth rate of the strange mode in the O4 model, which is independent of $k$, is calculated to be $\langle\omega_{\mathrm{ac}}\rangle \approx -0.37 / t_0$. Thus, the strange modes are, on average, damped in the region where the power spectra are computed. The convective growth rate, on the other hand, is on average positive but the average value is two orders of magnitude smaller than the result for the small-wavenumber mode in Fig. \ref{fig:density_power}. 
    %Since neither theoretical growth rate agrees well with the computed values from the simulation and since the $k$-dependence in the top panel of Fig. \ref{fig:rates_vs_k} does not show the characteristic $\sim 1/k^2$ or $\sim k^0$ behavior, it seems very unlikely that the structure formation in the O4 model can be primarily explained by the convective instability or the strange mode instability. 
    Since the theoretical convective growth rate does not agree with the computed growth rates and since the $k$-dependence in the top panel of Fig. \ref{fig:rates_vs_k} does not show the characteristic $\sim 1/k^2$ scaling, the convective instability is unlikely to be the origin of structure formation in the O4 model. While the plateau in the top panel of Fig. \ref{fig:rates_vs_k} could be in line with the $\sim k^0$ scaling of the strange mode instability, these strange modes are expected to be damped in the studied region and therefore, the strange mode instability also seems to be an unlikely explanation.

    %  \begin{figure}[t!]
    %     \centering
    %     \includegraphics[width=9cm]{ost_th1.png}
    %     \caption{Average density structure from the snapshot at $t = 1 t_{dyn}$ of the O4 model with respect to radial coordinate (bottom axis) or gas temperature (top axis). The red dashed vertical lines indicate the predicted region where the convective instability will operate and the blue ones indicate the predicted region where the strange mode instability will operate. The black dotted line indicates the photosphere.}
    %     \label{fig:ost_th1}
    % \end{figure}

    \subsection{Wolf-Rayet stars}

    In the bottom panel of Fig. \ref{fig:init}, we show the average density structure at $t = 2 t_0$ for the WR model, where we once again indicate the region in which the convective mode could be unstable according to Eq. (\ref{eq:stab_con}) by the red dashed lines. The convective region is located deeper below the photosphere as compared to the results for the O4 model and it also seems to extend further. The key difference with the previous O4 model is the strange mode region. Whereas in the O4 model, the strange mode instability region \fix{is} clearly bounded and located close to the photosphere, this mode is predicted to be overstable over the entire radial extent of the WR model, extending far down to the lower boundary. Due to the initial conditions of the WR model being less smooth than those of the O4 model, the theoretical growth rates fluctuate heavily within the convective region, even going negative at multiple %radial 
    vertical points. Therefore, the estimated mean convective growth rate for $k = 88 / R_\mathrm{c}$ over the range from $z = 1.3 R_{\mathrm{c}}$ to $z = 1.8 R_{\mathrm{c}}$ and $y=-0.25 R_\mathrm{c}$ to $y=0.25 R_\mathrm{c}$ (consistent with the slice that was used to produce the bottom panel in Fig. \ref{fig:density_power}) still comes out slightly damped with $\langle\omega_{\mathrm{g}}\rangle \approx -0.0050 / t_0$. The mean growth rate for the strange mode in this same region is $\langle\omega_{\mathrm{ac}} \rangle \approx 1.53 / t_{0}$. 
    %The strange mode growth rate is slightly larger than the values in the bottom panel of Fig. \ref{fig:rates_vs_k}, but it does match the order or magnitude of the computed growth rates quite well, especially on the large-wavenumber end. 
    The strange mode growth rate matches the order of magnitude of the computed growth rates in the bottom panel of Fig. \ref{fig:rates_vs_k} quite well, especially on the small-wavenumber end.
    However, the $k$-dependence in the bottom panel of Fig. \ref{fig:rates_vs_k} also does not match the characteristic $\sim 1/k^2$ or $\sim k^0$ behavior of either instability. As both the theoretical value of the convective growth rate as well as the $k$-dependence does not match the results of the WR model, the convective instability can most likely be excluded as a potential structure formation mechanism. For the strange modes, however, the result is less clear as the theoretical strange mode growth rate does match the computed growth rates quite well, but the $k$-dependence is different. Although, it must be noted that the theoretical $k$-dependence turns out to be different in the limit of fully negligible gas pressure, see Section \ref{sect:k-dep} for further discussion. 

    \subsection{Comparison}
    \label{sec:comp}
    The differences in theoretical growth rates and instability regions between the O4 model and the WR model can be explained by analyzing Eqs. (\ref{eq:con_timescales}) and (\ref{eq:ac_drag}), as done in Section \ref{lin_stability}, and giving some characteristic numbers at the peak of the iron bump in the respective models. \\
    \\
     For the convective mode, the growth rate is determined by the ratio of the diffusion timescale and the total Brunt-V\"ais\"al\"a frequency. For $k = 100 / R_\odot$, which corresponds to $k_\mathrm{WR} = 100 / R_\mathrm{c}$ and $k_\mathrm{O} = 1350 / R_\mathrm{c}$, we find that $t_{\mathrm{diff,O}} \approx 0.192$ s and $t_{\mathrm{diff,WR}} \approx 0.013$ s so the diffusion time in the O4 model is a factor fourteen larger than the diffusion time in the WR model.
     % We find for $k_O = 49 / R_{\mathrm{c}} \approx 4 / R_{\odot}$ and $k_{WR} = 8 / R_{\mathrm{c}} = 8 / R_{\odot}$ that $t_{diff,O} \approx 150 s$ and $t_{diff,WR} \approx 2 s$. 
     This difference can be explained by referring back to our earlier rewriting of $t_{\mathrm{diff}} = \tau_{\lambda}t_{\mathrm{ff}}$. 
     % The wavelengths differ by a factor of two, which affects both $t_{\mathrm{ff}}$ and $\tau_{\lambda}$. 
     As the wavenumber (and hence the wavelength) is the same, $t_\mathrm{ff}$ does not change between the two models, so the difference in diffusion time comes entirely from the factor $\tau_\lambda$. 
     %The opacity is a factor two larger in the O4 model and the density is a factor seven larger, leading to this factor fourteen difference in the optical depth.
     This difference in optical depth comes mainly from the larger density in the iron bump region of the O4 model. Thus, due to the larger optical depth, the O4 model \fix{experiences} less damping by radiative diffusion.
     % Furthermore, since the opacity is roughly the same in the O4 model and the WR model, the optical depth of the perturbation is a factor of ten larger in the O4 model because the density differs by a factor of ten compared to the WR model (see Fig. \ref{fig:init}). 
     %Thus, since the optical depth of the perturbation is larger in the O4 model, the diffusion time is larger as well, which means that the convective mode will experience less damping by radiative diffusion than in the WR model.
    The total Brunt-V\"ais\"al\"a frequency depends on gradients of density and (gas or radiation) pressure. 
    %For the O4 model, we find $N_{\mathrm{tot}}^2 \approx -1.9 \times 10^{-6}~\mathrm{s}^{-2} \approx -52 / t_0^2$ and for the WR model, we find $N_{\mathrm{tot}}^2 \approx 6 \times 10^{-4}~\mathrm{s}^{-2} \approx 409 / t_0^2$. 
    The density inversion in the initial conditions of the O4 model makes the second term in Eq. (\ref{eq:brunt_rad}) absolutely positive, which means that $N_{\mathrm{tot}}^2 < 0$ \fix{is} always true. Thus, the density inversion that arises from trying to maintain hydrostatic equilibrium when $\Gamma > 1$ is inherently unstable to convection in luminous stellar envelopes \citep{langer1997}.
    Since the WR star is able to launch a wind from the iron bump, there is no density inversion and therefore not necessarily an instability. This, together with the quick radiative diffusion, makes the convective instability less prominent in the WR stars.
    %Note that $N_{\mathrm{tot}}^2 < 0$ means that the mode is unstable. So, at the peak of the iron bump in the WR model, the convective mode is (on average) actually damped. It was to be expected that the O4 model would be more unstable than the WR model because of its density inversion. Indeed, a density inversion makes the second term in Eq. (\ref{eq:brunt_rad}) absolutely positive, which means that $N_{\mathrm{tot}}^2 < 0$ will always be true. Thus, the density inversion that arises from trying to maintain hydrostatic equilibrium when $\Gamma > 1$ is inherently unstable to convection in luminous stellar envelopes \citep{langer1997}. Since the WR star is able to launch a wind from the iron bump, there is no density inversion and therefore not necessarily an instability. This, together with the quick radiative diffusion, makes the convective instability less prominent in the WR stars.
    \\ \\
    Another major difference between the O4 model and the WR model is the strange mode growth rates and instability region. This comes from the interplay between the different terms in Eq. (\ref{eq:ac_drag}). 
    %As previously discussed, the factor in front of Eq. (\ref{eq:ac_drag}) depends on the Eddington factor $\Gamma$ and $g$, which means that this factor will be larger for the WR stars since they have higher luminosities (larger $\Gamma$) and smaller core radii (larger $g$). However, this is a scaling factor and it has no impact on determining the instability region. 
    The boundaries of the instability region are determined by the competition between $\Theta_{\rho}$ and $D_\mathrm{r}$. Interestingly, we find that $\Theta_{\rho}$ is on the same order in the O4 model and the WR model with $\Theta_{\rho} \approx 0.15$. Therefore, the difference between the two models comes entirely from the radiative drag term. We find that $D_\mathrm{r}$ in the O4 model is around a factor of seventy larger than in the WR model. 
    %While the gas sound speed is slightly larger in the O4 model,
    The main difference in this radiation drag term comes from the flux factor in the definition of $D_\mathrm{r}$, since the flux of the WR model is indeed larger than the flux in the O4 model.
    %by approximately a factor sixty. 
    We can also relate this to the local flux temperature $\Tilde{T}_{\mathrm{eff}}$ that we defined previously. This $\Tilde{T}_{\mathrm{eff}}$ is indeed larger in WR stars than in \fix{O stars} and this explains why WR stars experience less radiative drag. Due to the smaller radiative drag, we expect the strange modes to be more prominent in the WR model than in the O4 model.
    
    \section{Discussion}
    \label{discussion}

    \subsection{Growth rate $k$-dependence}
    \label{sect:k-dep}
    %\CV{Add something about Jiang et al. 2013}
    %\JS{Perhaps this one?  https://ui.adsabs.harvard.edu/abs/2013ApJ...763..102J/abstract}

    As discussed above, the top panel of Fig. \ref{fig:rates_vs_k} shows that the growth rates of both the O4 model and the WR model have a very peculiar $k$-dependence in the sense that they do not match either instability studied in Section \ref{lin_stability}. Therefore, we attempt to give suggestions for alternative formation mechanisms.\\
    \\
    An instability that has not been discussed here is the Rayleigh-Taylor (RT) instability which arises when a higher density gas overlays a lower density gas. However, the density inversion in Fig. \ref{fig:init} is inherently unstable to this instability as well. Furthermore, although the WR model does not show any density inversions on average, some outflowing density inversions may appear in the early stages of the simulation (see Fig. 3 in \citealt{moens2022}). Therefore, both the O4 model and the WR model could in principle experience RT-like instabilities, although we expect a larger effect in the O4 model due to the density inversion being quite prominent. Such a RT-like instability \fix{does} not appear in the stability analysis in Section \ref{lin_stability} because we assumed the background state to be locally uniform. To find these RT instabilities, we should have considered a background state with a density jump at an interface with certain boundary conditions \citep{jacquet2011}. This could, in principle, be done but this is outside the scope of the current work. \citet{jacquet2011} studied the linear growth of radiative RT instabilities in both the optically thin and the optically thick limit. Furthermore, in the appendix of \citet{jiang2013}, radiative RT instabilities in the limit of negligible absorption opacity were studied. In all of these different limits, the same $\omega \sim \sqrt{k}$ scaling of the classical RT instability was found.  
    Thus, although it is not entirely clear which limits can be safely applied to the simulations studied here, radiative RT instabilities generally seem to lead to the same $k$-dependence as found in our simulations.
    %In \citet{jacquet2011}, the linear growth of radiative RT instabilities was studied in both the optically thin and the optically thick limit. In the optically thick regime and the short wavelength limit, they find that the result reduces to the classical RT result where $\omega \sim \sqrt{k}$. This could potentially be an explanation for the dependence seen in Fig. \ref{fig:rates_vs_k}. However, \cite{jacquet2011} assume two semi-infinite media separated by an infinitely thin interface with a large density contrast which is very different from the continuous density inversion encountered here. Furthermore, the ratio between the peak of the density inversion and the lowest density point below is only a factor of three. As such, it is (at least formally) unclear whether the result by \citet{jacquet2011} holds for the current configuration even if the basic dependencies are the same. %\JS{From my quick look though, it seems like the Jiang study I refer to above is more similar to our set up. For example their appendix seems relevant, and they do find same k-dependence there.}
    \\
    \\
    Another possibility is the strange mode instability result in the negligible gas pressure limit found in \citet{bs03}, their section 3.3, in which they \fix{neglected} gas pressure altogether and \fix{found} 
    \begin{equation}
        \omega^2 = i \frac{\kappa}{c} \vec{k} \cdot \vec{F} \Theta_{\rho}. \label{eq:negligible_gas}
    \end{equation}
    This result also reproduces this $\omega \sim \sqrt{k}$ dependence.
    It could be argued that this limit applies to these systems, as they are radiation-pressure dominated, although the gas pressure still accounts for up to 30\% of the total pressure at the density inversion in the O4 model. In the WR model, however, on average only about 0.5\% of the total pressure is accounted for by the gas pressure. Thus, 
    %particularly 
    at least for the O4 model, it is unclear whether this limit can be applied. \\
    \\
    Therefore, both RT-like instabilities and strange mode instabilities in the limit of negligible gas pressure would produce the same scaling we find in our simulations. However, since these instabilities have the same $k$-dependence, it is difficult to distinguish between them and we have also not formally derived these scalings for our problem set-up. The task of conclusively identifying which instability is leading to this $\omega \sim \sqrt{k}$ scaling is left for future work.

    \subsection{Observational implications}

    \paragraph{Atmospheric micro and macroturbulence.} The turbulence generated in the envelopes and atmospheres of massive stars can lead to multiple observational effects. One noteworthy effect is the relation between turbulent motions and the so-called microturbulence or macroturbulence, additional 
    %velocity fields 
    ad-hoc velocity broadening terms
    that were introduced to explain excessive line broadening that could not be explained by rotational or thermal broadening \citep{conti1977,gray,simondiaz2017}. %\JS{some refs: Conti \& Ebbet (1977), Simon-Diaz et al. 2017, Gray (book on stellar photospheres)}
    %Traditionally, in spectroscopic work, microturbulence refers to isotropic Gaussian distributed turbulent motions which are optically thin, whereas macroturbulence concerns optically thick (and thus typically larger scale) motions. Additionally, the Gaussian distributions for such macroturbulence are sometimes assumed to be isotropic and sometimes to occur only in the radial or tangential (lateral) directions \citep{gray}.
    %\JS{you could write something like like: 'Traditionally, in spectroscopic work microturbulence refers to isotropic Gaussian distributed turbulent motions that are optically thin, whereas macroturbulence concerns optically thick (thus typically larger scale) motions. Additionally, the Gaussian distributions for such macroturbulence are sometimes assumed to be isotropic and sometimes to occur only in the radial and/or tangential (lateral) directions (Gray- 'stellar photospheres' book)'} 
    It has been suggested that the turbulent sub-surface layer associated with the iron bump can excite gravity waves which lead to velocity fluctuations at the surface and can therefore possibly be related to microturbulence \citep{cantiello}. Moreover, supersonic turbulence up to the surface, as seen in simulations like those presented here, is a natural explanation for the inferred macroturbulent velocities \citep{jiang2015,schultz2022,debnath2024}. On the other hand, gravity waves excited by core-convection or this sub-surface turbulent zone and traveling up to the stellar surface have also been discussed with respect to the origin of such macroturbulence \citep{aerts2009,simondiaz2017,serebriakova2024}. The origin of these structures \fix{is} important in the context of macroturbulence as the driving mechanism of the turbulent structures \fix{impacts}
    %their final velocities. 
    both their final velocities and their distributions. For example, gravity waves are excited preferentially in the lateral directions, which %would 
    may lead to turbulent velocity profiles dominated by the tangential component. However, such laterally dominated distributions are not necessarily the outcome for the excitation mechanisms %that seem to be favored by this work. 
    suggested in Section \ref{sect:k-dep}. Since different velocity distributions give rise to different shapes of the line profile broadening functions (see overview in \citealt{gray}), detailed high-resolution spectroscopy may be able to distinguish between these different scenarios.
    %Furthermore, as seen in Fig. \ref{fig:init}, different instabilities can lead to different instability regions which in turn determines the existence of a stable radiative layer on top of the turbulent iron bump region. The presence of such a stable radiative layer just below the surface determines the viability of a gravity wave scenario as proposed in, among others, \citet{cantiello}. \JS{Mmm, not sure about this last thing, at least not for simulations here. For lower luminosities and thus weaker turbulence, yes. Then this could be a viable transition from motions generated in this region vs. motions originating from the deeper core-regions.}\\
    \\
    
    \paragraph{Leakage of light.} As mentioned above, the origin of the structures also has an impact on their final shapes which may affect the porosity properties of the turbulent atmosphere. Such porosity generally refers to the additional light leakage that can occur from an inhomogeneous gaseous system as compared to a homogeneous one of the same mass, and is important in a wide range of astrophysical environments. 
    %(for example, light propagation through the intergalactic medium  in the early Universe, through the interstellar medium, and through the host galaxy of an active galactic nucleus). 
    In the context of stellar winds of massive stars, such a porous atmosphere can lead to reduced mass loss rates for the same total luminosity \citep{shaviv2000}. The quantitative amount of light-leakage can depend on the shapes of the structures. For example, finger-like, radially extended structures such as those seen in this paper may lead to larger escape fractions for photons emitted in the radial direction than in the lateral. As one specific example, this might have a pronounced effect on the number of continuum photons emitted from the deeper photosphere which are able to interact with resonant lines in the wind. Due to Doppler shifts, these interactions can only occur on surfaces of constant line-of-sight velocities and hence such "velocity-porosity" effects \citep{oskinova2007,sundqvist2010,Owocki_2014} have been shown to be particularly important for line formation in massive-star winds \citep{prinja2013,surlan2012,hawcroft2021,brands2022}. Current implementations of velocity-porosity into standard spectroscopic analysis tools \citep{sundqvist2018}, however, assume isotropic effects. In view of the results found here, it is thus not clear to what extent the amount of line photon leakage through the atmosphere is properly captured within current spectroscopic studies of massive stars.

    \subsection{On the nature of sub-surface turbulent zones in massive stars}

    It has long been known that stellar envelopes approaching the Eddington limit will trigger a sub-surface convection zone \citep{langer1997,maeder}. Due to the iron bump pushing part of the stellar envelope to the Eddington limit, the sub-surface convection zone that \fix{is} created is often linked to the near-surface turbulence in massive stars. Our results, however, suggest that convection is actually not the main formation mechanism of the envelope and atmospheric turbulence in \fix{O stars} and WR stars. \\
    \\
    It is important to note, though, that the initial conditions of the presented simulations may have an impact on the formation of structures. Especially for the O4 model, the initial conditions assume a hydrostatic envelope in radiative equilibrium, resulting in a density inversion. The final average structure of the non-linear simulations, however, no longer shows such a density inversion. Furthermore, the average structure also involves significant turbulent pressure support. Therefore, the average structure deviates quite heavily from the initial conditions, meaning that the relaxation of the initial conditions is significant in these simulations.
    %final average structure, so 
    %will be significant. 
    Further work is needed to study exactly how sensitive the structure formation is to the initial conditions. %For example, the same simulation could in principle be done using an average profile as initial conditions, thereby removing the density inversion, to analyze how this would influence the growth of structure.
    %the potential RT instabilities. 
    Nevertheless, with this in mind, our results here can still be used to gain further insights into the underlying physics of the formation of envelope turbulence in massive stars. \\
    \\
    We find that, at least in the case of \fix{O stars} and WR stars, while the convective instability is indeed expected to be present in the iron bump region, the initial structure formation does not show the $1/k^2$ dependence expected for radiative convection. This indicates that other instabilities, such as potentially the strange mode instability and RT-like instabilities, could be the dominant driving force behind this turbulence. Additionally, analysis of the final, non-linear turbulent states in simulations of \fix{O stars} and WR stars shows that energy transport by enthalpy (or, convection) is very inefficient in these atmospheres, with $\ga$ 90 \% of the total energy being carried by diffuse radiation even at the peak of the iron bump \citep{jiang2015,debnath2024}. By contrast, the atmospheric turbulent pressure support is very significant, that is, reaching similar levels as radiation pressure in the photospheric layers of the O4 model. Therefore, in view of our results, the treatment of this turbulent zone as a classical convection zone in 1D stellar structure and evolution models should be revisited. \\
    \\
    Interestingly, a similar deviation from a $\omega \sim 1/k^2$ dependence \fix{is} found in the StarTop model in \citet{jiang2015}. This is the model whose parameters most closely resemble the parameters of the O4 model. Similar to our O4 model, their StarTop model is in the regime where convection is an inefficient mode of energy transport. By contrast, in their model StarDeep, which lies in the efficient convection regime, they do approximately reproduce the $\sim 1/k^2$ dependence. This suggests that the convective instability might become more significant as enthalpy becomes more efficient in transporting energy. We notice a potential anti-correlation between such convective energy transport and the strength of the strange mode instability. Namely, efficiency of energy transport through convection \fix{scales} as $\sim c_i E/F_{\rm diff}$ \citep{grafener2012,jiang2015,debnath2024}, but this is precisely the damping term $D_\mathrm{r}$ (in the radiation-pressure dominated limit) that competes against the opacity driving term $\Theta_\rho$ in the strange mode instability. Thus, in regimes of efficient convective energy transport, the strange mode instability is expected to also be strongly damped (or even non-existent). Thus, it might be that the structure formation mechanism and the turbulent properties systematically change as one moves to different stellar regimes with varying $T_{\mathrm{eff}}$ (see also \citealt{schultz2023}).  \\
    \\
    
    \section{Summary and outlook}
    \label{conclusion}

    The turbulent sub-surface layers of \fix{O stars} and WR stars play an important role both for the stellar structure as well as for the emergent spectra. However, the origin of this turbulence, albeit often linked to a sub-surface convection zone triggered by the iron bump, has not been extensively studied. In this paper, we studied the formation mechanism of the emergent turbulence in multi-D simulations of WR stars and \fix{O stars} by \citet{moens2022} and \citet{debnath2024}, respectively. Based on work by \citet{bs03}, we identified multiple instabilities that could operate in the optically thick, radiation-pressure dominated envelopes of these stars. Two possible instabilities \fix{are} found: the convective instability and a local variant of the strange mode instability. Studying the analytical growth rates of these instabilities shows that the convective growth rate depends on wavenumber $k$ as $\omega_\mathrm{g} \sim 1/k^2$ and the strange mode is independent of wavenumber, that is, $\omega_\mathrm{a} \sim k^0$. The growth rates of the structures in the simulations were deduced by computing power spectra of the relative density, tracking the power evolution over time and fitting an exponential to the results. We \fix{find} that $\omega \sim \sqrt{k}$ for both the O4 model and the WR model. Therefore, the $k$-dependence from the simulations is incompatible with either instability studied here. Instead, this $\omega \sim \sqrt{k}$ scaling could potentially be compatible with the Rayleigh-Taylor instability \citep{jacquet2011,jiang2013} or the strange mode instability in the limit of negligible gas pressure \citep{bs03}. In any case, we clearly find that the structure formation in both models does not agree with the classical radiatively modified convection picture as we do not reproduce the expected $\omega \sim 1/k^2$ scaling. This is a significant result since, as previously mentioned, this sub-surface turbulent layer is often treated as a sub-surface convective layer (for instance, in stellar evolution codes such as MESA \citealt{paxton2013}) in the sense that energy transport is often assumed to be efficient, which is not the case for the simulations analyzed here. 
    %always the case \citep{jiang2015, debnath2024}. 
    Moreover, the different formation mechanism \fix{impacts} the final shape and size of the structures. \\
    \\
    As the exact structure formation mechanism in these simulations remains inconclusive for now, future work will attempt to identify other instabilities by modifying the stability analysis accordingly. Another direct follow-up to this work would be to quantify the properties of the emergent non-linear turbulent structures. In particular, we would like to quantify the energy distribution in the turbulent medium by computing energy power spectra and then comparing this to classical turbulence theory (see for example \citealt{schmidth2009}). It would also be useful to quantify the characteristic length scale of these structures and distinguish between the lateral and vertical components. \\
    Furthermore, we wish to extend this analysis to different stellar regimes such as Luminous Blue Variables or Red Supergiants. As discussed in the previous section, we expect to find differences between our results here and those regimes where convective energy transport is supposed to be more efficient. 

\begin{acknowledgements}

The computational resources used for this work were provided by Vlaams Supercomputer Centrum (VSC) funded by the Research Foundation-Flanders (FWO) and the Flemish Government. CVdS, JOS, DD and NM acknowledge the support of the European Research Council (ERC) Horizon Europe under grant agreement number 101044048. JOS further acknowledges the support of the Belgian Research Foundation Flanders (FWO) Odysseus program under grant number G0H9218N, and from FWO grant G077822N. The authors thank the referee Dr. Achim Feldmeier for constructive criticism on the manuscript. The authors would like to thank all members of the KUL-EQUATION group for fruitful discussion, comments, and suggestions. The following packages were used to analyze the data: {\fontfamily{qcr}\selectfont NumPy} \citep{harris_2020}, {\fontfamily{qcr}\selectfont SciPy} \citep{virtanen_2020}, {\fontfamily{qcr}\selectfont matplotlib} \citep{hunter_2007}, {\fontfamily{qcr}\selectfont Python amrvac\_reader} \citep{keppens_2020}. 

\end{acknowledgements}

\bibliographystyle{aa}
\bibliography{bibliography.bib} 

\begin{thebibliography}{67}
\expandafter\ifx\csname natexlab\endcsname\relax\def\natexlab#1{#1}\fi

\bibitem[{Aerts {et~al.}(2009)Aerts, Puls, Godart, \& Dupret}]{aerts2009}
Aerts, C., Puls, J., Godart, M., \& Dupret, M.-A. 2009, A\&A, 508, 409

\bibitem[{{Blaes} \& {Socrates}(2003)}]{bs03}
{Blaes}, O. \& {Socrates}, A. 2003, \apj, 596, 509

\bibitem[{Brands {et~al.}(2022)Brands, de~Koter, Bestenlehner, Crowther, Sundqvist, Puls, Caballero-Nieves, Abdul-Masih, Driessen, Garc{\'\i}a, {et~al.}}]{brands2022}
Brands, S.~A., de~Koter, A., Bestenlehner, J.~M., {et~al.} 2022, A\&A, 663, A36

\bibitem[{{Cantiello} {et~al.}(2009){Cantiello}, {Langer}, {Brott}, {de Koter}, {Shore}, {Vink}, {Voegler}, {Lennon}, \& {Yoon}}]{cantiello}
{Cantiello}, M., {Langer}, N., {Brott}, I., {et~al.} 2009, A\&A, 499, 279

\bibitem[{{Castor} {et~al.}(1975){Castor}, {Abbott}, \& {Klein}}]{castor1975}
{Castor}, J.~I., {Abbott}, D.~C., \& {Klein}, R.~I. 1975, \apj, 195, 157

\bibitem[{{Conti} \& {Ebbets}(1977)}]{conti1977}
{Conti}, P.~S. \& {Ebbets}, D. 1977, \apj, 213, 438

\bibitem[{{Crowther}(2007)}]{crowther2007}
{Crowther}, P.~A. 2007, \araa, 45, 177

\bibitem[{{Debnath} {et~al.}(2024){Debnath}, {Sundqvist}, {Moens}, {Van der Sijpt}, {Verhamme}, \& {Poniatowski}}]{debnath2024}
{Debnath}, D., {Sundqvist}, J.~O., {Moens}, N., {et~al.} 2024, \aap, 684, A177

\bibitem[{{Dessart} \& {Owocki}(2002)}]{dessart2002}
{Dessart}, L. \& {Owocki}, S.~P. 2002, \aap, 393, 991

\bibitem[{{Eversberg} {et~al.}(1998){Eversberg}, {L{\'e}pine}, \& {Moffat}}]{eversberg1998}
{Eversberg}, T., {L{\'e}pine}, S., \& {Moffat}, A. F.~J. 1998, \apj, 494, 799

\bibitem[{{Gautschy} \& {Glatzel}(1990)}]{gautschyandglatzel}
{Gautschy}, A. \& {Glatzel}, W. 1990, \mnras, 245, 597

\bibitem[{{Glatzel}(1994)}]{glatzel1994}
{Glatzel}, W. 1994, \mnras, 271, 66

\bibitem[{{González-Torà} {et~al.}(2024){González-Torà}, {Sander}, {Sundqvist}, {Debnath}, {Delbroek}, {Josiek}, {Lefever}, {Moens}, {Van der Sijpt}, \& {Verhamme}}]{gonzalez2024}
{González-Torà}, G., {Sander}, A.~A.~C., {Sundqvist}, J.~O., {et~al.} 2024, \aap, submitted

\bibitem[{{Gr{\"a}fener} {et~al.}(2002){Gr{\"a}fener}, {Koesterke}, \& {Hamann}}]{grafener2002}
{Gr{\"a}fener}, G., {Koesterke}, L., \& {Hamann}, W.~R. 2002, \aap, 387, 244

\bibitem[{{Gr{\"a}fener} {et~al.}(2012){Gr{\"a}fener}, {Owocki}, \& {Vink}}]{grafener2012}
{Gr{\"a}fener}, G., {Owocki}, S.~P., \& {Vink}, J.~S. 2012, \aap, 538, A40

\bibitem[{{Gray}(2008)}]{gray}
{Gray}, D.~F. 2008, {The Observation and Analysis of Stellar Photospheres}

\bibitem[{{Hamann} \& {Gr{\"a}fener}(2004)}]{hamann2004}
{Hamann}, W.~R. \& {Gr{\"a}fener}, G. 2004, \aap, 427, 697

\bibitem[{{Harris} {et~al.}(2020){Harris}, {Millman}, {van der Walt}, {Gommers}, {Virtanen}, {Cournapeau}, {Wieser}, {Taylor}, {Berg}, {Smith}, {Kern}, {Picus}, {Hoyer}, {van Kerkwijk}, {Brett}, {Haldane}, {del R{\'\i}o}, {Wiebe}, {Peterson}, {G{\'e}rard-Marchant}, {Sheppard}, {Reddy}, {Weckesser}, {Abbasi}, {Gohlke}, \& {Oliphant}}]{harris_2020}
{Harris}, C.~R., {Millman}, K.~J., {van der Walt}, S.~J., {et~al.} 2020, \nat, 585, 357

\bibitem[{Hawcroft {et~al.}(2021)Hawcroft, Sana, Mahy, Sundqvist, Abdul-Masih, Bouret, Brands, de~Koter, Driessen, \& Puls}]{hawcroft2021}
Hawcroft, C., Sana, H., Mahy, L., {et~al.} 2021, A\&A, 655, A67

\bibitem[{{Hillier} \& {Lanz}(2001)}]{hillier2001}
{Hillier}, D.~J. \& {Lanz}, T. 2001, in Astronomical Society of the Pacific Conference Series, Vol. 247, Spectroscopic Challenges of Photoionized Plasmas, ed. G.~{Ferland} \& D.~W. {Savin}, 343

\bibitem[{{Hillier} \& {Miller}(1998)}]{hillier1998}
{Hillier}, D.~J. \& {Miller}, D.~L. 1998, \apj, 496, 407

\bibitem[{{Howarth} {et~al.}(1997){Howarth}, {Siebert}, {Hussain}, \& {Prinja}}]{howarth1997}
{Howarth}, I.~D., {Siebert}, K.~W., {Hussain}, G. A.~J., \& {Prinja}, R.~K. 1997, \mnras, 284, 265

\bibitem[{{Hunter}(2007)}]{hunter_2007}
{Hunter}, J.~D. 2007, Computing in Science and Engineering, 9, 90

\bibitem[{{Iglesias} \& {Rogers}(1996)}]{opal}
{Iglesias}, C.~A. \& {Rogers}, F.~J. 1996, \apj, 464, 943

\bibitem[{{Jacquet} \& {Krumholz}(2011)}]{jacquet2011}
{Jacquet}, E. \& {Krumholz}, M.~R. 2011, \apj, 730, 116

\bibitem[{{Jiang} {et~al.}(2015){Jiang}, {Cantiello}, {Bildsten}, {Quataert}, \& {Blaes}}]{jiang2015}
{Jiang}, Y.-F., {Cantiello}, M., {Bildsten}, L., {Quataert}, E., \& {Blaes}, O. 2015, \apj, 813, 74

\bibitem[{{Jiang} {et~al.}(2013){Jiang}, {Davis}, \& {Stone}}]{jiang2013}
{Jiang}, Y.-F., {Davis}, S.~W., \& {Stone}, J.~M. 2013, \apj, 763, 102

\bibitem[{{Keppens} {et~al.}(2023){Keppens}, {Popescu Braileanu}, {Zhou}, {Ruan}, {Xia}, {Guo}, {Claes}, \& {Bacchini}}]{keppens2023}
{Keppens}, R., {Popescu Braileanu}, B., {Zhou}, Y., {et~al.} 2023, \aap, 673, A66

\bibitem[{Keppens {et~al.}(2021)Keppens, Teunissen, Xia, \& Porth}]{keppens_2020}
Keppens, R., Teunissen, J., Xia, C., \& Porth, O. 2021, Computers \& Mathematics with Applications, 81, 316

\bibitem[{Langer(1997)}]{langer1997}
Langer, N. 1997, in Luminous Blue Variables: Massive Stars in Transition, Vol. 120, 83

\bibitem[{{Lepine}(1994)}]{lepine1994}
{Lepine}, S. 1994, \apss, 221, 371

\bibitem[{{Levermore} \& {Pomraning}(1981)}]{pomraning}
{Levermore}, C.~D. \& {Pomraning}, G.~C. 1981, \apj, 248, 321

\bibitem[{Maeder(2009)}]{maeder}
Maeder, A. 2009, Physics, Formation and Evolution of Rotating Stars, Astronomy and Astrophysics Library (Berlin, Heidelberg: Springer Berlin Heidelberg)

\bibitem[{{Michaux} {et~al.}(2014){Michaux}, {Moffat}, {Chen{\'e}}, \& {St-Louis}}]{michaux2014}
{Michaux}, Y. J.~L., {Moffat}, A. F.~J., {Chen{\'e}}, A.-N., \& {St-Louis}, N. 2014, \mnras, 440, 2

\bibitem[{{Moens} {et~al.}(2022b){Moens}, {Poniatowski}, {Hennicker}, {Sundqvist}, {El Mellah}, \& {Kee}}]{moens2022}
{Moens}, N., {Poniatowski}, L.~G., {Hennicker}, L., {et~al.} 2022b, \aap, 665, A42

\bibitem[{{Moens} {et~al.}(2022a){Moens}, {Sundqvist}, {El Mellah}, {Poniatowski}, {Teunissen}, \& {Keppens}}]{moens2021}
{Moens}, N., {Sundqvist}, J.~O., {El Mellah}, I., {et~al.} 2022a, A\&A, 657, A81

\bibitem[{{Moffat} {et~al.}(1988){Moffat}, {Drissen}, {Lamontagne}, \& {Robert}}]{moffat1988}
{Moffat}, A. F.~J., {Drissen}, L., {Lamontagne}, R., \& {Robert}, C. 1988, \apj, 334, 1038

\bibitem[{Oskinova {et~al.}(2007)Oskinova, Hamann, \& Feldmeier}]{oskinova2007}
Oskinova, L., Hamann, W.-R., \& Feldmeier, A. 2007, A\&A, 476, 1331

\bibitem[{Owocki(2014)}]{Owocki_2014}
Owocki, S.~P. 2014, Instabilities in the Envelopes and Winds of Very Massive Stars (Springer International Publishing), 113–156

\bibitem[{{Owocki} \& {Puls}(1996)}]{owocki1996}
{Owocki}, S.~P. \& {Puls}, J. 1996, \apj, 462, 894

\bibitem[{{Owocki} \& {Rybicki}(1984)}]{owocki1984}
{Owocki}, S.~P. \& {Rybicki}, G.~B. 1984, \apj, 284, 337

\bibitem[{{Owocki} \& {Rybicki}(1985)}]{owocki1985}
{Owocki}, S.~P. \& {Rybicki}, G.~B. 1985, \apj, 299, 265

\bibitem[{{Owocki} \& {Rybicki}(1991)}]{owocki1991}
{Owocki}, S.~P. \& {Rybicki}, G.~B. 1991, \apj, 368, 261

\bibitem[{{Paxton} {et~al.}(2013){Paxton}, {Cantiello}, {Arras}, {Bildsten}, {Brown}, {Dotter}, {Mankovich}, {Montgomery}, {Stello}, {Timmes}, \& {Townsend}}]{paxton2013}
{Paxton}, B., {Cantiello}, M., {Arras}, P., {et~al.} 2013, \apjs, 208, 4

\bibitem[{{Poniatowski} {et~al.}(2021){Poniatowski}, {Sundqvist}, {Kee}, {Owocki}, {Marchant}, {Decin}, {de Koter}, {Mahy}, \& {Sana}}]{poniatowski2021}
{Poniatowski}, L.~G., {Sundqvist}, J.~O., {Kee}, N.~D., {et~al.} 2021, A\&A, 647, A151

\bibitem[{Prinja \& Massa(2013)}]{prinja2013}
Prinja, R. \& Massa, D. 2013, A\&A, 559, A15

\bibitem[{{Puls} {et~al.}(2006){Puls}, {Markova}, {Scuderi}, {Stanghellini}, {Taranova}, {Burnley}, \& {Howarth}}]{puls2006}
{Puls}, J., {Markova}, N., {Scuderi}, S., {et~al.} 2006, \aap, 454, 625

\bibitem[{{Puls} {et~al.}(2020){Puls}, {Najarro}, {Sundqvist}, \& {Sen}}]{puls2020}
{Puls}, J., {Najarro}, F., {Sundqvist}, J.~O., \& {Sen}, K. 2020, \aap, 642, A172

\bibitem[{{Puls} {et~al.}(2005){Puls}, {Urbaneja}, {Venero}, {Repolust}, {Springmann}, {Jokuthy}, \& {Mokiem}}]{puls2005}
{Puls}, J., {Urbaneja}, M.~A., {Venero}, R., {et~al.} 2005, \aap, 435, 669

\bibitem[{{Rubio-D{\'\i}ez} {et~al.}(2022){Rubio-D{\'\i}ez}, {Sundqvist}, {Najarro}, {Traficante}, {Puls}, {Calzoletti}, \& {Figer}}]{rubiodiez2021}
{Rubio-D{\'\i}ez}, M.~M., {Sundqvist}, J.~O., {Najarro}, F., {et~al.} 2022, \aap, 658, A61

\bibitem[{{Rybicki} {et~al.}(1990){Rybicki}, {Owocki}, \& {Castor}}]{rybicki1990}
{Rybicki}, G.~B., {Owocki}, S.~P., \& {Castor}, J.~I. 1990, \apj, 349, 274

\bibitem[{{Saio} {et~al.}(1998){Saio}, {Baker}, \& {Gautschy}}]{saio}
{Saio}, H., {Baker}, N.~H., \& {Gautschy}, A. 1998, \mnras, 294, 622

\bibitem[{{Sander} {et~al.}(2012){Sander}, {Hamann}, \& {Todt}}]{sander2012}
{Sander}, A., {Hamann}, W.~R., \& {Todt}, H. 2012, \aap, 540, A144

\bibitem[{{Santolaya-Rey} {et~al.}(1997){Santolaya-Rey}, {Puls}, \& {Herrero}}]{santolaya1997}
{Santolaya-Rey}, A.~E., {Puls}, J., \& {Herrero}, A. 1997, \aap, 323, 488

\bibitem[{{Schmidt} {et~al.}(2009){Schmidt}, {Federrath}, {Hupp}, {Kern}, \& {Niemeyer}}]{schmidth2009}
{Schmidt}, W., {Federrath}, C., {Hupp}, M., {Kern}, S., \& {Niemeyer}, J.~C. 2009, \aap, 494, 127

\bibitem[{{Schultz} {et~al.}(2022){Schultz}, {Bildsten}, \& {Jiang}}]{schultz2022}
{Schultz}, W.~C., {Bildsten}, L., \& {Jiang}, Y.-F. 2022, \apjl, 924, L11

\bibitem[{{Schultz} {et~al.}(2023){Schultz}, {Bildsten}, \& {Jiang}}]{schultz2023}
{Schultz}, W.~C., {Bildsten}, L., \& {Jiang}, Y.-F. 2023, \apjl, 951, L42

\bibitem[{{Serebriakova} {et~al.}(2024){Serebriakova}, {Tkachenko}, \& {Aerts}}]{serebriakova2024}
{Serebriakova}, N., {Tkachenko}, A., \& {Aerts}, C. 2024, \aap, 692, A245

\bibitem[{{Shaviv}(2000)}]{shaviv2000}
{Shaviv}, N.~J. 2000, \apjl, 532, L137

\bibitem[{{Shaviv}(2001)}]{shaviv2001}
{Shaviv}, N.~J. 2001, \apj, 549, 1093

\bibitem[{Sim{\'o}n-D{\'\i}az {et~al.}(2017)Sim{\'o}n-D{\'\i}az, Godart, Castro, Herrero, Aerts, Puls, Telting, \& Grassitelli}]{simondiaz2017}
Sim{\'o}n-D{\'\i}az, S., Godart, M., Castro, N., {et~al.} 2017, A\&A, 597, A22

\bibitem[{Sundqvist \& Puls(2018)}]{sundqvist2018}
Sundqvist, J. \& Puls, J. 2018, A\&A, 619, A59

\bibitem[{Sundqvist {et~al.}(2010)Sundqvist, Puls, \& Feldmeier}]{sundqvist2010}
Sundqvist, J., Puls, J., \& Feldmeier, A. 2010, A\&A, 510, A11

\bibitem[{{Sundqvist} {et~al.}(2018){Sundqvist}, {Owocki}, \& {Puls}}]{sunqvist2018b}
{Sundqvist}, J.~O., {Owocki}, S.~P., \& {Puls}, J. 2018, \aap, 611, A17

\bibitem[{{\v{S}}urlan {et~al.}(2012){\v{S}}urlan, Hamann, Kub{\'a}t, Oskinova, \& Feldmeier}]{surlan2012}
{\v{S}}urlan, B., Hamann, W.-R., Kub{\'a}t, J., Oskinova, L.~M., \& Feldmeier, A. 2012, A\&A, 541, A37

\bibitem[{{Virtanen} {et~al.}(2020){Virtanen}, {Gommers}, {Oliphant}, {Haberland}, {Reddy}, {Cournapeau}, {Burovski}, {Peterson}, {Weckesser}, {Bright}, {van der Walt}, {Brett}, {Wilson}, {Millman}, {Mayorov}, {Nelson}, {Jones}, {Kern}, {Larson}, {Carey}, {Polat}, {Feng}, {Moore}, {VanderPlas}, {Laxalde}, {Perktold}, {Cimrman}, {Henriksen}, {Quintero}, {Harris}, {Archibald}, {Ribeiro}, {Pedregosa}, {van Mulbregt}, \& {SciPy 1. 0 Contributors}}]{virtanen_2020}
{Virtanen}, P., {Gommers}, R., {Oliphant}, T.~E., {et~al.} 2020, Nature Methods, 17, 261

\bibitem[{{Xia} {et~al.}(2018){Xia}, {Teunissen}, {El Mellah}, {Chan{\'e}}, \& {Keppens}}]{xia2018}
{Xia}, C., {Teunissen}, J., {El Mellah}, I., {Chan{\'e}}, E., \& {Keppens}, R. 2018, \apjs, 234, 30

\end{thebibliography}

\begin{appendix}

\section{Additional considerations for a mean background outflow}
\label{sec:app}
    \subsection{Advection timescale}

    \label{sec:outflow}
    As previously mentioned, WR stars can launch an optically thick wind from the sub-surface layers whereas an \fix{O star} has a fully optically thin wind, initiated outside the photosphere, on top of a turbulent deeper atmosphere. For WR stars, the average sonic point can therefore lie within the iron bump region, meaning that there might already be a significant outflow in the structure formation region. This leads to a competing timescale: any structures that grow due to an instability must be able to grow to non-linear scales before the wind expels a significant amount of mass from the unstable region. This same reasoning was already put forward in \citet{cantiello} for the helium opacity bump, which is located even closer to the photosphere. 
    %\JS{A bit too much jargon to just say 'helium opacity bump' here. Introduce what this actually physically represents.} 
    As in \citet{cantiello}, we may define a certain mass loss timescale for our WR model as the time $t_{\Dot{M}}$ it takes for the wind to expel an amount of mass equal to the mass in the unstable region, namely
    \begin{equation}
        t_{\Dot{M}} = \Delta M / \Dot{M},
    \end{equation}
    where $\Delta M$ is the total mass contained within the unstable region and $\Dot{M}$ is the mass loss rate. We focus on the convective region indicated in Fig. \ref{fig:init} which extends over $\Delta z \approx 1.3R_{\mathrm{c}}$. 
    We calculate a characteristic $\Delta M = 4 \pi z^2 \Delta z \langle \rho \rangle = 1.94 \times 10^{-9} ~M_\odot$ in a spherical shell at a distance $z$, where we have taken the 
    %radius 
    $z$-coordinate at the midpoint of the convective region so $z=1.95R_\mathrm{c}$ and $\langle \rho \rangle = 1.9 \times 10^{-10} ~\mathrm{g}/\mathrm{cm}^{3}$ is the laterally averaged density at this midpoint. The typical mass loss rate in this convective region can be calculated as the %local mass flux times the surface area a spherical star would have at that vertical distance. That is, equating $z = r$, $\Dot{M} = 4 \pi r^2\langle \varv_r \rho \rangle = 6.5 \times 10^{-5} M_\odot / \mathrm{yr}$
    mass loss rate at the midpoint $\Dot{M} = 4 \pi z^2\langle \varv_z \rho \rangle = 6.5 \times 10^{-5} ~M_\odot / \mathrm{yr}$. 
    % In this convective region, the outflow has an average density $\Bar{\rho} \approx 4 \cdot 10^{-10} g/cm^{3}$ at the midpoint. We calculate $\Delta M = \Bar{\rho} \Delta V$, with $\Delta V = 0.65 R_{\odot}^3$, and we find $\Delta M \approx 4 \cdot 10^{-11} M_{\odot}$. We determine the mass-loss rate as the average mass flux through the surface area $A$, so $\Dot{M} = \Bar{\rho {\varv}_r} A$, 
    % %\JS{so really this should be average over rho*v, not the terms individually. Not sure if it makes difference here though (but generally < rho*v > ne < rho > < v>)?} 
    % where $A = 0.5 R_{\odot}^2$, and we find $\Dot{M} \approx 1.1 \cdot 10^{-6} M_{\odot} / year$. \JS{I must have asked you this before, sorry, but do we know why this mdot estimate is so low here?} 
    Using these estimates, we find
    \begin{equation}
        t_{\Dot{M}} \approx 940 \text{ s}.
    \end{equation}
    This time is equivalent to the typical advection time of a fluid parcel across the convective region.
    %\JS{You should mention that this in the end is same/similar as considering a typical advection time across the region t =v/del r (which, imo, is actually a better and neater way of viewing this).} 
    This time needs to be compared to the typical time it takes for convection to develop. We can estimate the minimal time that is needed for a perturbation to grow due to the convective instability. The maximum growth rate in the convective region of the WR stars is $\omega_\mathrm{g} = 0.34 / t_0$ so the minimal time that is needed for a perturbation to grow by a factor of $e$ is $t = 1 / \omega_\mathrm{g} \approx 2400$ s. This time is already larger than $t_{\Dot{M}}$ and it is important to note that this estimate is the minimal convective growth time, which \fix{is} only valid at a specific point in the convective region, but not everywhere. Thus, on average, it \fix{takes} much longer for convection to develop which means that the mass loss time $t_{\Dot{M}}$ \fix{is} significantly smaller than this growth time. Therefore, the convective region is completely cleared of material before convection can fully develop. \\
    \\
    The case for the strange modes is different. Firstly, the instability region is larger than the convective region. Considering only the sub-photospheric layers, the extent of the instability region is now $\Delta z \approx 2.2 R_{\mathrm{c}}$ which results in $\Delta M = 3.3 \times 10^{-9} M_{\odot}$ and $\Dot{M} = 6.5 \times 10^{-5} M_\odot / \mathrm{yr}$. Therefore, the advection time becomes 
    \begin{equation}
        t_{\Dot{M}} \approx 1600 \text{ s}.
    \end{equation}
    The maximum acoustic growth rate in the sub-surface region is $\omega \approx 1.91 / t_0$ which results in a minimal growth time $t \approx 430$ s. Thus, the growth time is now significantly shorter than the advection time. Furthermore, these strange modes are propagating waves as opposed to the gravity waves. Hence, the strange mode instability is a traveling instability instead of an instability at a fixed 
    %radial 
    vertical position. Therefore, the strange modes will not be impacted by material being cleared out of a certain fixed region as they will propagate out of this region also. For these reasons, we do not expect the growth of the strange modes to be substantially impacted by the sub-surface outflow.

    \subsection{Eulerian vs Lagrangian frame}
    
\begin{figure}[t!]
        \centering
        \includegraphics[width=9.2cm]{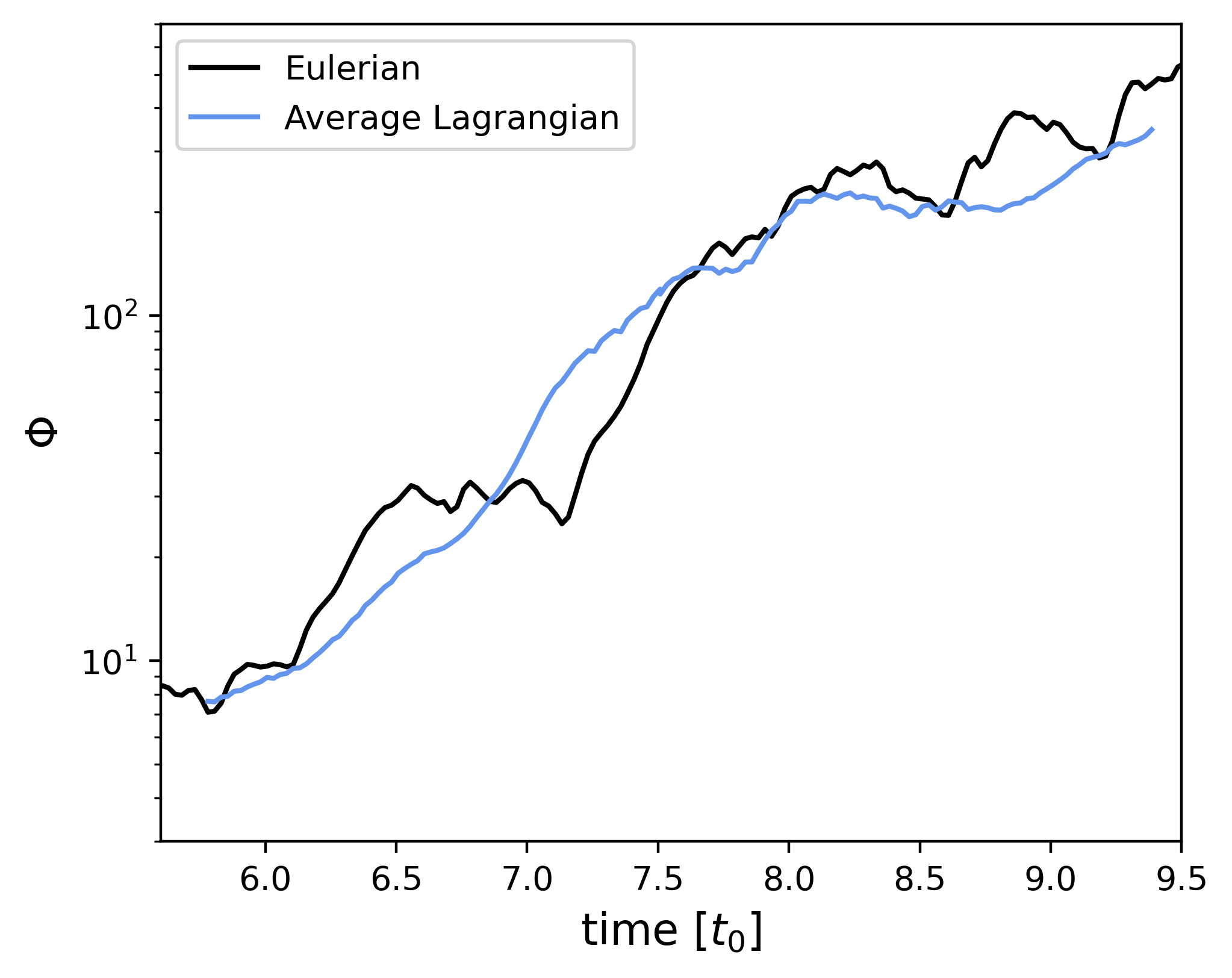}
        \caption{Power evolution comparison between the Eulerian frame and the Lagrangian frame. The black curve shows the power evolution in the usual Eulerian frame and the blue curve shows the power evolution for the average Lagrangian frame, where the averaging was done for four different starting positions (see text).}
        \label{fig:movingbox_avg}
    \end{figure}

\begin{figure}[t!]
        \centering
        \includegraphics[width=9cm]{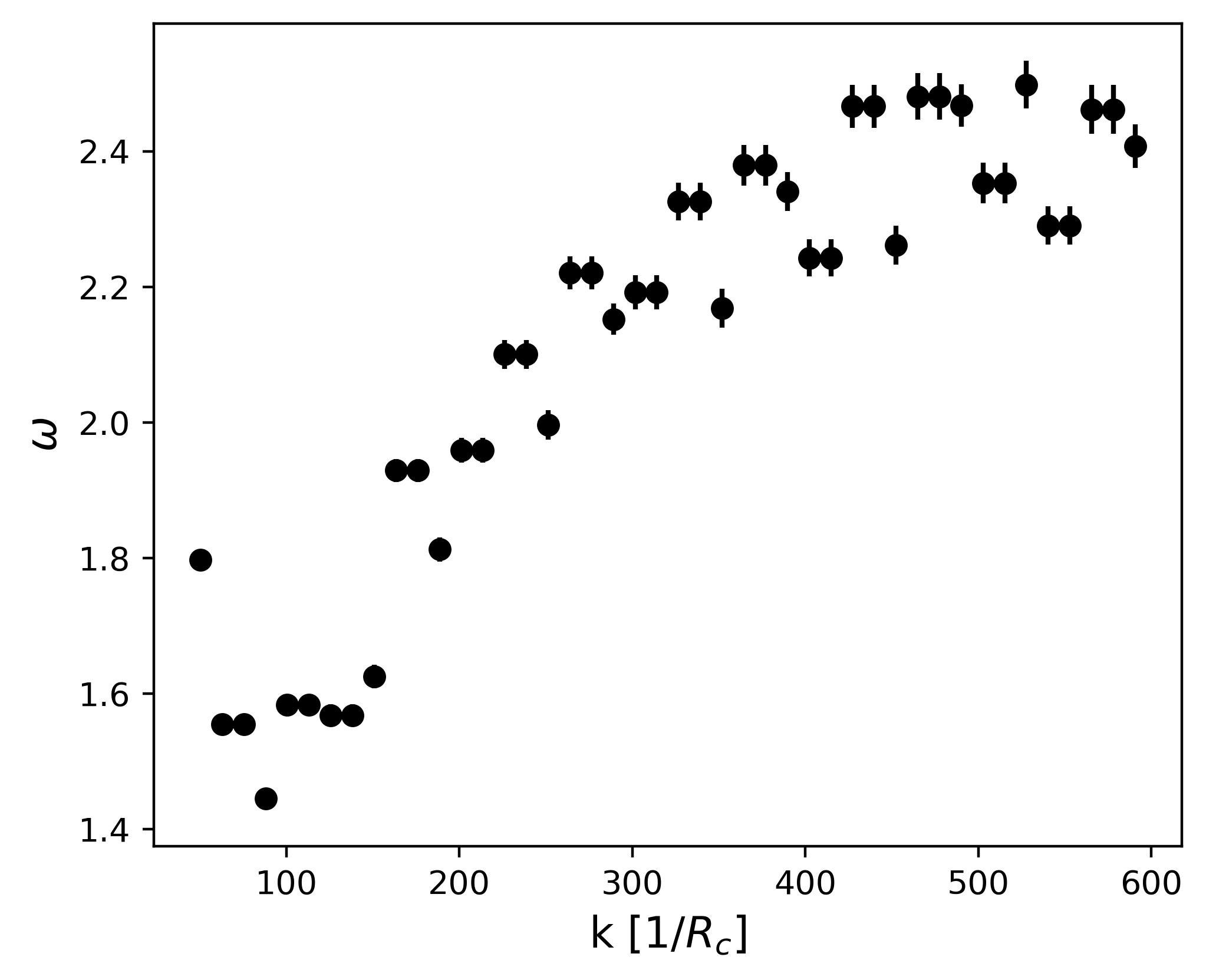}
        \caption{Average growth rates for multiple wavenumbers from the Lagrangian frame analysis. The final growth rate for a certain wavenumber is a weighted average of thirty growth rates from different starting positions between $z = 1.1 R_c$ and $z = 1.4 R_c$ in steps of $\Delta z = 0.01 R_c$.}
        \label{fig:avg_rates}
    \end{figure}

    As explained in Section \ref{lin_stability}, the resulting dispersion relation from the linear stability analysis is obtained by assuming perturbations on top of a certain background equilibrium state. This equilibrium state was taken to be static, but this approximation is questionable for the WR stars, in particular, as the velocities in the region of interest are already significant. For the \fix{O stars}, this approximation seems reasonable since the average velocity in the sub-surface layers is close to zero. The power spectra in Section \ref{results} were computed by taking a cut of the grid at a fixed location (see Section \ref{results}), that is, by studying the structures in an Eulerian reference frame. However, in reality, the result for the dispersion relation in Section \ref{lin_stability} is only valid in a reference frame moving with the local fluid velocity, that is, a Lagrangian frame. This means that the cut of the grid should actually follow the local fluid velocity. Implementing this in practice is non-trivial since the gas is accelerating so the top part of the cut will have a larger average gas velocity than the bottom part in which case the size of the cut would not be conserved. To simplify, we \fix{started} with a cut of $0.5 R_{\mathrm{c}} \times 0.5 R_{\mathrm{c}}$, as before, and at each time step, we \fix{calculated} the total average radial gas velocity $\varv_{\mathrm{avg}}$ within this cut. Then, for the next time step, we \fix{moved} the cut by a 
    %radial 
    vertical distance $\varv_{\mathrm{avg}} \Delta t$. For each time step, the power in a certain mode \fix{was} calculated using the procedure outlined in Section \ref{results}. To avoid complications with the different refinement levels as the cut moves through the box, we re-ran the WR simulation on a full grid at the base refinement of $1024 \times 128$ cells. In this method, the initial position of the cut is a free parameter. The only restriction we want to impose is that the cut moves through the full iron bump region, meaning that the initial location of the lower boundary of the cut should be below the iron bump. Therefore, we \fix{sampled} four different initial lower boundary positions, namely at $z = 1.1 R_{\mathrm{c}}$, $z = 1.2 R_{\mathrm{c}}$, $z = 1.3 R_{\mathrm{c}}$ and $z = 1.4 R_{\mathrm{c}}$ to study how this initial position might affect the final curves. Inspection of the results shows that the initial position does affect the computed growth rates and we believe this is due to the time-dependent nature of these simulations, meaning that not all regions are always equally structured and by changing the initial position, we \fix{track} more or less structured regions with different behavior. As such, we \fix{took} an average of the four different starting positions in order to obtain a more general view of the power evolution obtained from this Lagrangian method. In Fig. \ref{fig:movingbox_avg}, this average Lagrangian power evolution is shown as the blue curve and the black curve shows the evolution curve in the Eulerian frame as used above. From this figure, it is clear that in an average sense, the power evolution is qualitatively the same between the Lagrangian method and the Eulerian method. As a further comparison, we show the average growth rates from this Lagrangian frame analysis as a function of wavenumber in Fig. \ref{fig:avg_rates}. The average \fix{was} computed as a weighted average of the results from thirty different starting positions, ranging from $z = 1.1 R_c$ to $z = 1.4 R_c$ in steps of $\Delta z = 0.01 R_c$. This figure shows that the general $k$-dependence agrees well between the Eulerian and the Lagrangian analysis. Repeating the previous analysis using the Lagrangian method comes with some complications. Firstly, as already mentioned, the results are quite sensitive to the starting position. Secondly, as we are only interested in the optically thick envelope, the analysis can only be done as long as the cut stays below the photosphere. Because of this, we can only track the power evolution for shorter periods of time. Due to these reasons, and because the two methods seem to give qualitatively similar results, the Eulerian analysis \fix{is} favored in this work.
    

\end{appendix}

\end{document}